%% file: vldb-veristrong-arxiv.tex
\documentclass[sigconf, nonacm]{acmart}

\newcommand\vldbdoi{XX.XX/XXX.XX}
\newcommand\vldbpages{XXX-XXX}
\newcommand\vldbvolume{1}
\newcommand\vldbissue{1}
\newcommand\vldbyear{2026}
\newcommand\vldbauthors{\authors}
\newcommand\vldbtitle{\shorttitle}
\newcommand\vldbavailabilityurl{https://github.com/CzxingcHen/VeriStrong}
\newcommand\vldbpagestyle{plain}

\input{setup/packages}

\input{setup/macro}

\begin{document}

\title{Fast Verification of Strong Database Isolation\\
	(Extended Version)
}

\author{Zhiheng Cai}
\affiliation{%
	\institution{Tsinghua University}
	\country{}
}
\email{cai-zh24@mails.tsinghua.edu.cn}

\author{Si Liu}
\affiliation{%
	\institution{ETH Zurich}
	\country{}
}
\email{si.liu@inf.ethz.ch}

\author{Hengfeng Wei}
\affiliation{%
	\institution{Hunan University}
	\country{}
}
\authornote{The corresponding author who is with
	College of Computer Science and Electronic Engineering,
	Hunan University.
	The initial draft of this work was prepared
	while Hengfeng Wei and Zhiheng Cai were with
	Software Institute, Nanjing University.}
\email{hfwei@hnu.edu.cn}

\author{Yuxing Chen}
\affiliation{%
	\institution{Tencent Inc.}
	\country{}
}
\email{axingguchen@tencent.com}

\author{Anqun Pan}
\affiliation{%
	\institution{Tencent Inc.}
	\country{}
}
\email{aaronpan@tencent.com}

\input{sections/abs}

\maketitle

\pagestyle{\vldbpagestyle}
\begingroup\small\noindent\raggedright\textbf{PVLDB Reference Format:}\\
\vldbauthors. \vldbtitle. PVLDB, \vldbvolume(\vldbissue): \vldbpages, \vldbyear.\\
\href{https://doi.org/\vldbdoi}{doi:\vldbdoi}
\endgroup
\begingroup
\renewcommand\thefootnote{}\footnote{\noindent
This work is licensed under the Creative Commons BY-NC-ND 4.0 International License. Visit \url{https://creativecommons.org/licenses/by-nc-nd/4.0/} to view a copy of this license. For any use beyond those covered by this license, obtain permission by emailing \href{mailto:info@vldb.org}{info@vldb.org}. Copyright is held by the owner/author(s). Publication rights licensed to the VLDB Endowment. \\
\raggedright Proceedings of the VLDB Endowment, Vol. \vldbvolume, No. \vldbissue\ %
ISSN 2150-8097. \\
\href{https://doi.org/\vldbdoi}{doi:\vldbdoi} \\
}\addtocounter{footnote}{-1}\endgroup

\ifdefempty{\vldbavailabilityurl}{}{
\vspace{.3cm}
\begingroup\small\noindent\raggedright\textbf{PVLDB Artifact Availability:}\\
The source code, data, and/or other artifacts have been made available at \url{\vldbavailabilityurl}. 
\endgroup
}

\input{sections/intro-nobi-0423}


\input{sections/prelim-nobi-0424}
\input{sections/epg-nobi-0424}
\input{sections/alg-nobi-0424}

\input{sections/eval-nobi-0424}

\input{sections/related-nobi-0424}

\input{sections/discussion-nobi-0424}

\section*{Acknowledgments}
We appreciate the anonymous reviewers for their valuable feedback.
Hengfeng Wei was supported by the NSFC (62472214).
Si Liu was supported by an ETH Zurich Career Seed Award.

\bibliographystyle{ACM-Reference-Format}
\bibliography{ref}

\clearpage
\appendix

\input{sections/si-nobi-0424}
\input{appendix/app-si}

\input{appendix/app-proof}
\input{appendix/cdclt}
\input{appendix/app-algo}
\input{appendix/app-correctness}
\input{appendix/app-eval}

\input{appendix/app-enea}
\input{appendix/app-notation-table}

\end{document}

%% file: setup/packages.tex
\usepackage{amsmath, amsfonts}
\usepackage{amssymb}
\usepackage{graphicx}
\usepackage{textcomp}
\usepackage[dvipsnames]{xcolor}
\usepackage{subcaption}
\usepackage{xspace}
\usepackage{tikz} 
\usepackage{bbding} 
\usepackage{diagbox} 
\usepackage[normalem]{ulem}

\usepackage{algorithm}
\usepackage{algpseudocodex}
\usepackage{algorithmicx}

\usepackage{environ}
\usepackage{adjustbox}

\usepackage{pgfplots}
\usetikzlibrary{patterns}
\usepgfplotslibrary{groupplots}

\usepackage{todonotes}

\usepackage{enumitem}

\usepackage[most]{tcolorbox}

\usepackage{wrapfig}

\usepackage{amsthm}  

\usetikzlibrary{spy}

\usepackage{tabularx}
\usepackage{soul}

\usepackage{multirow}

\usepackage{capt-of}

\usepackage{mathpartir}
\usepackage{mathtools}

\usepackage{marginnote}

%% file: setup/macro.tex
\newcommand{\nobi}[1]{{\color{orange}{#1}}}
\newcommand{\zhiheng}[1]{{\textcolor{blue!75!black}{[zhiheng: #1]}}}

\newcommand{\ourtool}{\textsc{VeriStrong}\xspace}

\newcommand{\inlsec}[1]{\smallskip\noindent\textbf{#1.}}

\newcommand{\circled}[1]{{\textcircled{\footnotesize #1}}}

\newcommand{\rc}{\textsc{Read Committed}\xspace} 
\newcommand{\ra}{\textsc{Read Atomic}\xspace}

\newcommand{\si}{\textsc{Snapshot Isolation}\xspace}
\newcommand{\ser}{\textsc{Serializability}\xspace}

\newcommand{\red}[1]{\textcolor{red}{#1}}

\newcommand{\blue}[1]{\textcolor{blue}{#1}}

\newcommand{\brown}[1]{\textcolor{brown}{#1}}

\definecolor{ps1green}{HTML}{B7E6A5} 
\definecolor{ps2green}{HTML}{7CCBA2}
\definecolor{ps3green}{HTML}{46AEA0}
\definecolor{ps4green}{HTML}{089099}
\definecolor{ps5green}{HTML}{00718B}
\definecolor{ps6green}{HTML}{045275}
\definecolor{ps7green}{HTML}{003147}

\definecolor{diverge1}{HTML}{045275}
\definecolor{diverge2}{HTML}{089099}
\definecolor{diverge3}{HTML}{EF8733}
\definecolor{diverge4}{HTML}{FCDE9C}
\definecolor{diverge5}{HTML}{F0746E}
\definecolor{diverge6}{HTML}{DC3977}
\definecolor{diverge7}{HTML}{7C1D6F}

\definecolor{equalsoneoptcolor}{RGB}{253, 243, 208}
\definecolor{bigandsmallcyclesoptcolor}{RGB}{217, 232, 214}
\definecolor{otheroptcolor}{RGB}{251, 231, 207}

\newcommand{\true}{\textsf{true}}
\newcommand{\false}{\textsf{false}}
\newcommand{\formula}{\mathcal{F}}

\newcommand{\bigo}{\mathcal{O}} 

\newcommand{\F}{\mathcal{F}} 
\newcommand{\set}[1]{\{#1\}}

\newcommand{\Bset}[1]{\Big\{#1\Big\}}

\newcommand{\tuple}[1]{\langle#1\rangle} 
\newcommand{\btuple}[1]{\big\{#1\big\}}

\newcommand{\type}{\sf Type}
\newcommand{\typevar}{{\sf T}}

\newcommand{\notmodelssymbol}{\mathsf{S_{\#}}}

\newcommand{\uniqueval}{\textsf{UniqueValue}}

\newcommand{\duplicateval}{\textsf{DuplicateValue}}

\newcommand{\Key}{{\sf Key}}

\newcommand{\keyxvar}{\mathit{x}}
\newcommand{\keyyvar}{\mathit{y}}
\newcommand{\keyzvar}{\mathit{z}}
\newcommand{\Val}{{\sf Val}}
\newcommand{\valvar}{\mathit{v}}

\newcommand{\opid}{\iota}
\newcommand{\OpId}{\mathsf{OpId}}

\newcommand{\h}{\mathcal{H}}

\newcommand{\opset}{\mathit{O}}

\newcommand{\Op}{{\sf Op}}

\newcommand{\readevent}{{\sf R}}

\newcommand{\ReadTx}{{\sf ReadTx}}
\newcommand{\writeevent}{{\sf W}}

\newcommand{\WriteTx}{{\sf WriteTx}}

\newcommand{\appendevent}{{\sf A}}

\newcommand{\T}{\mathcal{T}}
\providecommand{\G}{}
\renewcommand{\G}{\mathcal{G}}
\renewcommand{\H}{\mathcal{H}}
\renewcommand{\P}{\mathcal{P}}
\newcommand{\V}{\mathcal{V}}
\newcommand{\E}{\mathcal{E}}
\newcommand{\M}{\mathcal{M}}

\newcommand{\SO}{\textsf{\textup{SO}}}
\newcommand{\WR}{\textsf{\textup{WR}}}
\newcommand{\WW}{\textsf{\textup{WW}}}
\newcommand{\RW}{\textsf{\textup{RW}}}

\newcommand{\CO}{\textsf{\textup{CO}}}

\newcommand{\acyclicencoding}{\textup{\sf Acyclic}}

\newcommand{\polysi}{PolySI\xspace}
\newcommand{\cobra}{Cobra\xspace}
\newcommand{\viper}{Viper\xspace}
\newcommand{\dbcop}{dbcop\xspace}
\newcommand{\minisat}{MiniSat\xspace} 

\newcommand{\rel}[1]{\xrightarrow{#1}}

\newcommand{\comp}{\;;\;}
\newcommand{\po}{{\sf po}}
\newcommand{\intaxiom}{\textsc{Int}}

\newcommand{\cardinality}[1]{\lvert #1 \rvert}
\newcommand{\totalsize}[1]{\lVert #1 \rVert}

\newcommand{\eithervar}{\mathit{either}}
\newcommand{\orvar}{\mathit{or}}

\newcommand{\eitheredges}{\mathit{either\_edges}}
\newcommand{\oredges}{\mathit{or\_edges}}
\newcommand{\wrvar}{\mathit{wr}}
\newcommand{\wredges}{\mathit{wr\_edges}}

\newcommand{\hist}{\H}

\newcommand{\C}{\mathcal{C}}

\newcommand{\wwcons}{\C^{\WW}}
\newcommand{\wrcons}{\C^{\WR}}
\newcommand{\cocons}{\C^{\CO}}
\newcommand{\consvar}{C}
\newcommand{\ccons}{\mathit{cons}}

\newcommand{\creachability}{\mathcal{R}}

\newcommand{\csovar}{\text{SO}}
\newcommand{\cwwvar}{\text{WW}}
\newcommand{\cwrvar}[1]{\text{WR}^{#1}}

\newcommand{\crwvar}{\text{RW}}
\newcommand{\cvars}{\mathit{Vars}}
\newcommand{\cclauses}{\mathit{Clauses}}

\newcommand{\cedges}{\mathit{edges}}

\newcommand{\clevel}[1]{\mathit{level}(#1)}

\newcommand{\ccycle}{\mathit{cycle}}
\newcommand{\cconflictcl}{\mathit{conflict\_clause}}
\newcommand{\clit}{\mathit{l}}

\newcommand{\cdclt}{\textsc{Cdcl(T)}}
\newcommand{\cdclth}{\textsc{Cdcl(T)}^{\textbf{H}}}

\newcommand{\sat}{\textsf{SAT}}
\newcommand{\unsat}{\textsf{UNSAT}}

\newcommand{\ug}{G^{*}} 
\newcommand{\eug}{E^{*}} 
\newcommand{\pa}{\mathcal{M}} 
\newcommand{\reasonvar}{\mathit{reason}}
\newcommand{\fromvar}{\mathit{from}}
\newcommand{\tovar}{\mathit{to}}

\makeatletter
\newsavebox{\measure@tikzpicture}
\NewEnviron{scaletikzpicturetowidth}[1]{%
  \def\tikz@width{#1}%
  \def\tikzscale{1}\begin{lrbox}{\measure@tikzpicture}%
  \BODY
  \end{lrbox}%
  \pgfmathparse{#1/\wd\measure@tikzpicture}%
  \edef\tikzscale{\pgfmathresult}%
  \BODY
}
\makeatother

\newcommand{\hStatex}[0]{\vspace{4pt}}

\newcommand{\statebox}[2][red]{%
  {}\tikz[baseline=(char.base)]{
    \node[
      draw=#1,
      fill=none,
      text=#1,
      font=\sffamily\bfseries\large,
      rounded corners=2pt,
      minimum width=0.32cm,
      minimum height=0.32cm,
      align=center,
      inner sep=0pt
    ] (char) {#2};
  }{}%
}
\colorlet{sgray}{gray!80!black}
\colorlet{sgreen}{green!40!black}
\colorlet{sred}{red!70!black}
\colorlet{sbrown}{brown!80!black}
\colorlet{sblue}{blue!50!black}

\newenvironment{proofsketch}{%
\renewcommand{\proofname}{Proof Sketch}\proof}{\endproof}

%% file: sections/abs.tex
\begin{abstract}
Strong isolation guarantees, such as serializability and snapshot isolation, are essential for maintaining data consistency and integrity in modern databases.
Verifying whether a database upholds its claimed guarantees is increasingly critical, as these guarantees form a contract between the vendor and its users. 
However, this task is   challenging, particularly in black-box settings, 
where only observable system behavior is available and often involves uncertain  dependencies between transactions.

In this paper, we present \ourtool, a fast verifier for strong database isolation. At its core is a novel formalism called hyper-polygraphs, which compactly captures  both certain and uncertain
transactional dependencies  in database executions.
Leveraging this formalism, we develop sound and complete encodings for verifying both serializability and snapshot isolation.
To achieve high efficiency, \ourtool tailors SMT solving to the characteristics of database workloads, in contrast to prior general-purpose approaches.
Our extensive evaluation across diverse benchmarks shows that 
\ourtool not only significantly outperforms state-of-the-art verifiers on the workloads they support, 
but also scales to large, general workloads beyond their reach, while maintaining high accuracy in detecting isolation anomalies.
\end{abstract}

%% file: sections/intro-nobi-0423.tex








\section{Introduction}

Strong isolation levels or guarantees---the I in ACID~\cite{TIS:Book2002}---such as \ser and \si, are essential for ensuring data  consistency and integrity in modern  databases.
These guarantees prevent subtle concurrency anomalies, such as lost updates, 
that can lead to incorrect application behavior. 
Once a database is deployed, its promised isolation guarantees become a binding contract between the vendor and its users. This raises a critical question: 
\emph{How can we be sure that the database  upholds its promise?}

Answering this question is nontrivial. 
Modern databases  have large codebases that are often inaccessible or too complex to reason about, even when available.
 This makes full verification of  isolation guarantees practically infeasible. 
 Black-box verification~\cite{Cobra:OSDI2020,Complexity:OOPSLA2019} offers an effective alternative: a verifier collects execution histories of database transactions as an \emph{external observer} and checks whether these histories satisfy the  isolation level in question.

However, verifying database histories remains  challenging. 
The verification problem for strong isolation levels has been proven to be NP-hard, even for verifying a single history~\cite{Complexity:OOPSLA2019}. 
This complexity arises from the fact that, as an external observer of a black-box database, a verifier must \emph{infer} the internal execution order of transactions, e.g., determining which transaction wrote an earlier version and which one wrote a later one. 
This challenge is further compounded by practical constraints: given limited time and memory, only a finite number of ``guesses'' can be made by the verifier, and thus only a  finite number of histories can be verified. 
Yet, examining more histories is highly desirable, either to increase the likelihood  of uncovering  anomalies or to strengthen confidence in their absence.

Recent years have seen significant progress in accelerating the verification of strong isolation guarantees~\cite{Cobra:OSDI2020,PolySI:VLDB2023,Viper:EuroSys2023,Elle:VLDB2020,Complexity:OOPSLA2019,MiniTransactions:ICDE2025}.
Representative verifiers include 
Cobra~\cite{Cobra:OSDI2020} for \ser,
PolySI~\cite{PolySI:VLDB2023}  and Viper~\cite{Viper:EuroSys2023} for \si,
and Elle~\cite{Elle:VLDB2020} 
for both   (a detailed discussion is provided in Section~\ref{section:related}).
Despite these advances, verification efficiency remains a key limitation.
Many verifiers have attempted to reduce verification overhead by leveraging advanced SMT (Satisfiability Modulo Theories) techniques~\cite{MonoSAT:AAAI2015}. 
However, 
general-purpose SMT solvers 
are prone to 
 missing optimization opportunities specific to 
database verification.

Beyond this, 
 two additional challenges make efficient verification even harder. 
First, all existing verifiers assume that transaction histories contain \emph{unique write values}, i.e.,  each read can be deterministically matched to a single write.
This  greatly simplifies the task of inferring the internal execution order
and serves as a key enabler of Elle's design in particular.
Yet this assumption does not hold in many real-world applications, where
different transactions often write the same value.
	For instance, in an e-commerce platform, multiple users may submit identical orders for the same quantity of an item; 
	in social media, thousands of users may ``like'' the same post. 
Likewise, standard database benchmarks that closely reflect real-world  workloads, such as Twitter~\cite{CTwitter} and RUBiS~\cite{CRUBis}, include \emph{duplicate write values} by default.
In fact, in our black-box setting, a (cloud) database  is unaware that it is under test; we act purely as an external  monitor, collecting the execution results of real workloads processed by the system.
Hence, supporting such general workloads is highly desirable and essential for practical applicability.

Moreover,
achieving  both soundness (no false alarms) and completeness (no missed bugs) without compromising efficiency remains challenging  in verifying strong isolation. 
In particular, 
identifying all anomalies in a history is computationally expensive, yet critical:  overlooking even a single anomaly can require significant effort to rediscover.
	This challenge is further exacerbated by recent findings~\cite{MariaDB-Bug,MySQL-Bug}, which reveal that certain anomalies arise \emph{only} when duplicate write values are present---precisely the scenarios exemplified in real-world workloads above.
Hence,
  relying on existing verifiers with unique write values risks both false negatives and false  positives,
   undermining the reliability of verification results.

\inlsec{Our Solution}
We present a  novel black-box verifier,  \ourtool, that checks database  histories against the two most widely used strong isolation guarantees, i.e., \ser and \si. 
\ourtool tackles the aforementioned challenges through an end-to-end verification pipeline. This includes
(i) representing and encoding general database histories,
(ii) establishing a sound and complete correspondence between the encoding and the verification algorithm,
and (iii) optimizing checking efficiency using  SMT techniques tailored for strong isolation verification.

At the core of our solution is a new formalism called \emph{hyper-polygraphs}, which captures both certain transactional dependencies (e.g., session order) and uncertain ones (e.g., version order). 
In black-box settings, uncertainties can also arise from ambiguous read-from relations caused by duplicate write values---an aspect that existing formalisms, such as polygraphs~\cite{SER:JACM1979,TIS:Book2002} and their variants~\cite{Cobra:OSDI2020,Viper:EuroSys2023,PolySI:VLDB2023}, cannot adequately express. 
Hyper-polygraphs offer two key advantages. 
First, they provide a compact and expressive means of encoding transactional dependencies, making them particularly amenable to SMT-based verification. 
Second, they are generic, capable of representing a wide range of dependency-based isolation  theories~\cite{Adya:PhDThesis1999,Complexity:OOPSLA2019,AnalysingSI:JACM2018}. 
In this paper,
we focus on Adya's theory~\cite{Adya:PhDThesis1999}, with another application discussed in Section~\ref{section:discuss}.

Building on this formalism, we develop sound and complete encodings for verifying both strong isolation guarantees, 
providing a rigorous theoretical foundation for reliable SMT-based verification. 
This enables our verifier to accurately (re)discover isolation anomalies in production databases (e.g., MariaDB), particularly those arising from duplicate write values that prior tools fail to handle.

To make verification fast,
our key insight is that leveraging workload-specific knowledge can greatly boost SMT solving efficiency. 
To this end, we propose two domain-specific optimizations. 
First, we offload part of the solver's effort to a lightweight preprocessing phase by proactively resolving small conflict patterns. 
This   reduces expensive backtracking during solving by eliminating many infeasible search paths early and 
 strikes a practical balance between preprocessing and solving costs,   improving overall performance.


Second, we design a polarity picking strategy---i.e., deciding whether to explore a dependency---guided by a dynamically maintained pseudo-topological order during SMT solving. 
Unlike general-purpose solvers that make such decisions ``blindly'', we leverage known partial orders, such as session order and certain read-from edges, to align these decisions with the likely transaction schedule enforced by the database.
This alignment steers the solver away from conflict-prone search paths, thereby enhancing efficiency.

\inlsec{Contributions}
Overall, we make the following contributions.
\begin{enumerate}[leftmargin=20pt]
	\item We introduce hyper-polygraphs, a new formalism that compactly characterizes  isolation levels, including \ser and \si, 
	over general database histories.

	\item We formally establish the soundness and completeness of our new characterizations, grounded in Adya's theory, providing a theoretical foundation for reliable SMT-based verification.
	
	\item We develop novel isolation verification algorithms by tailoring SMT solving to the characteristics of database workloads,
	incorporating optimizations that reduce solver overhead.

	\item We implement our approach in  \ourtool and  assess it extensively on a variety of benchmarks.
Results  show that \ourtool not only substantially outperforms state-of-the-art verifiers on the workloads they support, but also scales to large,  general workloads beyond their reach, while maintaining high accuracy in detecting isolation anomalies.
\end{enumerate}

This work  complements recent advances in  isolation verification along two dimensions. 
First, it complements prior efforts~\cite{Plume:OOPSLA2024,AWDIT:PLDI2025} on 
black-box verification of weaker levels like \rc, which are known to be less complex to verify~\cite{Complexity:OOPSLA2019}.
Second, it targets deployed databases, complementing recent work on verifying isolation in their designs~\cite{VerIso:VLDB2025,Maude:Liu2019} and implementations~\cite{Separation:ICFP2025}.

We focus on \ser as the running example throughout the paper, unless noted otherwise.
We defer to  Appendix~\ref{section:appendix-si} how our approach extends to verifying \si.

%% file: sections/prelim-nobi-0424.tex
\section{Background} \label{section:pre}

\input{sections/black-box-checking}

\input{sections/notations}
\input{sections/history}
\input{sections/ser-depgraph}


%% file: sections/black-box-checking.tex

\begin{figure}
	\centering
	\includegraphics[width=\columnwidth]{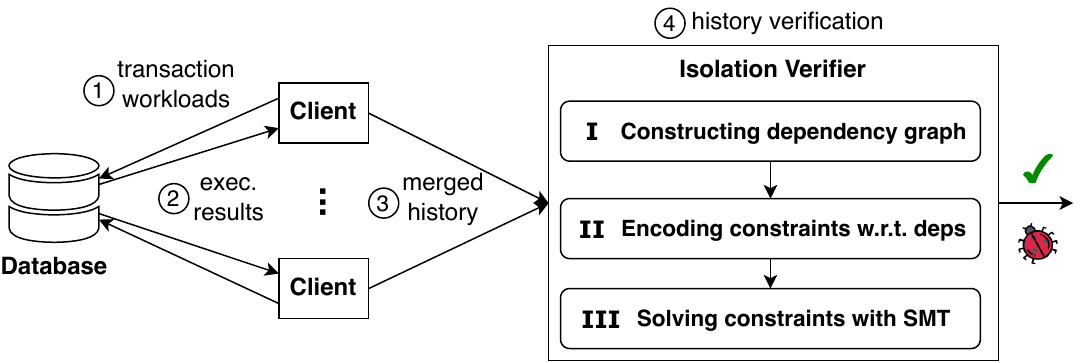}
	\captionsetup{skip=4pt}
	\caption{Workflow for verifying isolation guarantees in deployed databases.
		Steps II and III are specific to SMT-based approaches (see also Sections~\ref{section:baseline} and \ref{sec:alg}).
		}
	\label{fig:black-box-arch}
	\vspace{-2ex}
\end{figure}

Black-box verification of database isolation guarantees involves four key steps, as illustrated in Figure~\ref{fig:black-box-arch}. 
In Step \circled{1}, clients issue transactional requests to the database.
In Step \circled{2}, each client records the corresponding execution results, including the values read or written and the status of each transaction (either committed or aborted).
Next, in Step \circled{3}, the logs from all clients are merged into a single history, which is then passed to an isolation verifier. 
Finally, in Step \circled{4},  the verifier analyzes the history to determine whether it satisfies the specified isolation guarantee.

How does the verifier make this decision?
It constructs a dependency graph  based on, e.g., Adya's theory~\cite{Adya:PhDThesis1999} that is 
the \textit{de facto} formalization of isolation guarantees. 
The verifier then searches the graph for cycles, which indicate violations of guarantees such as \ser. 
Next, we 
recall    the formal definitions    underlying the history verification steps I--III from~\cite{PolySI:VLDB2023}.




%% file: sections/notations.tex



\inlsec{Relations}
A binary relation $R$ over a given set $A$ is a subset of $A \times A$, i.e., $R \subseteq A \times A$.
For $a, b \in A$, we use $(a, b) \in R$ and $a \rel{R} b$ interchangeably.
We use $R{?}$ and $R^{+}$ to denote the reflexive closure and the transitive closure of $R$, respectively.
A relation $R \subseteq A \times A$ is \emph{acyclic} if $R^{+} \cap I_{A} = \emptyset$,
where $I_{A} \triangleq \set{(a, a) \mid a \in A}$ is the identity relation on $A$.
Given two binary relations $R$ and $S$ over the set $A$, we define their composition as
$R \comp S = \{ (a, c) \mid \exists b \in A. a \rel{R} b \rel{S} c\}$.
A strict partial order is an irreflexive and transitive relation.
A strict total order is a relation that is a strict partial order and total.

%% file: sections/history.tex


\inlsec{Transactions}
We consider a distributed key-value store
that manages a set of keys $\Key = \set{\keyxvar, \keyyvar, \keyzvar, \dots}$;
each key is associated with a value from a set $\Val$.
The set of  operations, denoted by $\Op$, consists of both read and write operations:
$\Op = \set{\readevent_{\opid}(\keyxvar, \valvar), \writeevent_{\opid}(\keyxvar, \valvar)
  \mid \opid \in \OpId, \keyxvar \in \Key, \valvar \in \Val}$,
where $\OpId$ is the set of operation identifiers.
For simplicity, operation identifiers may be omitted.
We use $\_$ to represent an irrelevant  value in an operation,  e.g., $\readevent(\keyxvar, \_)$, which is implicitly existentially quantified.

Clients interact with the key-value store by issuing transactions.

\begin{definition} \label{def:transaction}
  A \emph{transaction} is a pair $(\opset, \po)$,
  where $\opset \subseteq \Op$ is a finite, non-empty set of operations;
   $\po \subseteq \opset \times \opset$ is a strict total order, referred to as the \emph{program order}, which indicates the execution order of operations within the  transaction.
\end{definition}

For a transaction $T$, we write $T \vdash \writeevent(\keyxvar, \valvar)$
if $T$ writes to  the key $\keyxvar$ and $\valvar$ is the last written value,
and  $T \vdash \readevent(\keyxvar, \valvar)$
if $T$ reads from $\keyxvar$ before writing to it,
and $\valvar$ is the value returned by the first such read.
We also define $\WriteTx_{\keyxvar} = \set{T \mid T \vdash \writeevent(\keyxvar, \_)}$  to denote the set of transactions that write to  $\keyxvar$.

\inlsec{Histories}
Transactions are grouped into client \emph{sessions}, where each session consists of a sequence of transactions.
We use \emph{histories} to record the client-visible outcomes of these transactions.

\begin{definition} \label{def:history}
  A \emph{history} is a pair $\h = (\T, \SO)$,
  where $\T$ is a set of transactions with disjoint sets of operations;
  $\SO \subseteq \T \times \T$ is the \emph{session order}, which is a union of strict total orders over disjoint sets of $\T$, each corresponding to  a distinct client session.
\end{definition}

 In line with existing formal models~\cite{AnalysingSI:JACM2018,Adya:PhDThesis1999,Complexity:OOPSLA2019},
we consider only committed transactions in a history; aborted transactions are handled
separately.
We also assume that every history contains a special transaction $T_{\bot}$
that writes the initial values of all keys.\footnote{In practice, we use multiple short, write-only transactions to initially populate the database. These transactions can be viewed as a single, logical write-only transaction.}
This transaction precedes all the other transactions in  $\SO$ across sessions.
Note that existing work commonly assumes \uniqueval{} histories, where each write to the same key assigns a distinct value, whereas real-world scenarios often involve \duplicateval{} writes.

%% file: sections/ser-depgraph.tex



\inlsec{Dependency Graphs}
A dependency graph extends a history by introducing three types of relations (or edges), $\WR$, $\WW$, and $\RW$, each capturing a  different kind of dependency between transactions.
These relations underlies the  formalization of isolation guarantees  in the style of Adya~\cite{AnalysingSI:JACM2018,Adya:PhDThesis1999}.
The $\WR$ relation connects a transaction that reads a value to the transaction that wrote it. The $\WW$ relation defines a strict total order, also referred to as the \emph{version order}~\cite{Adya:PhDThesis1999}, among transactions that write to the same key.
The $\RW$ relation is derived from $\WR$ and $\WW$, associating a transaction that reads a value to the one that  overwrites it based on the version order.

\begin{definition} \label{def:depgraph}
  A \emph{dependency graph} is a tuple $\G = (\T, \SO, \WR,$ $\WW, \RW)$,
  where $(\T, \SO)$ is a history and
  \begin{itemize}[leftmargin=15pt]
		\item $\WR: \Key \to 2^{\T \times \T}$ is such that
			\begin{itemize}[leftmargin=15pt]
				\item[--] $\forall \keyxvar \in \Key.\; \forall S \in \T.\;
					S \vdash \readevent(\keyxvar, \_) \!\!\implies\!\! \exists! T \in \T.\; T \rel{\WR(\keyxvar)} S$
				\item[--] $\forall \keyxvar \in \Key.\; \forall T, S \in \T.\;
					T \rel{\WR(x)} S \implies T \neq S \land
						\exists \valvar \in \Val.\; T \vdash \writeevent(\keyxvar, \valvar) \land S \vdash \readevent(\keyxvar, \valvar)$.
			\end{itemize}
		\item $\WW: \Key \to 2^{\T \times \T}$ is such that
		$\forall \keyxvar \in \Key$, $\WW(\keyxvar)$ is a strict total order on  $\WriteTx_{\keyxvar}$.
		\item $\RW: \Key \to 2^{\T \times \T}$ is such that
			$\forall T, S \in \T.\; \forall \keyxvar \in \Key.\;
				T \rel{\RW(\keyxvar)} S \iff T \neq S \land \exists T' \in \T.\; T' \rel{\WR(\keyxvar)} T \land T' \rel{\WW(\keyxvar)} S$.
  \end{itemize}
\end{definition}

We use $\exists!$ to denote unique existence.
For any component of a dependency graph $\G$, such as $\WW$, we write it as $\WW_{\G}$.
When the key $\keyxvar$  in $T \rel{R(\keyxvar)} S$ is irrelevant or clear from context, we write $T \rel{R} S$, where $R \in \set{\WR, \WW, \RW}$.

\inlsec{Characterizing Serializability}
\ser is characterized by the \emph{existence} of an acyclic dependency
graph, along with the internal consistency of each transaction, as
axiomatized by $\intaxiom$~\cite{AnalysingSI:JACM2018}.
The emphasis on existence is crucial: as external observers of a black-box database, verifiers cannot directly observe the internal execution order, e.g., $\WW$ between writes. Instead, they must infer a possible internal schedule. The existence of an acyclic dependency graph provides a witness that explains how the database could have scheduled the transactions in a way that satisfies \ser.

In addition,
 the $\intaxiom$  axiom ensures that a read on a key returns the same value as the most recent proceeding access---either a write or a read---to that key within the same transaction.

\begin{theorem} 
	\label{thm:ser-depgraph}
	For a history \emph{$\H = (\T, \SO)$},
	\emph{\begin{align*}
		\H \models\,\, &  \textsc{SER}  \iff \H \models \intaxiom \;\land \\
			&\exists \WR, \WW, \RW.\; \G = (\H, \WR, \WW, \RW) \;\land \\
			&\quad ((\SO_{\G} \cup \WR_{\G} \cup \WW_{\G} \cup \RW_{\G}) \text{\it\; is acyclic}).
	\end{align*}}
\end{theorem}

\begin{figure}[t]
	\begin{subfigure}[b]{0.45\columnwidth}
		\centering
		\includegraphics[width=\textwidth]{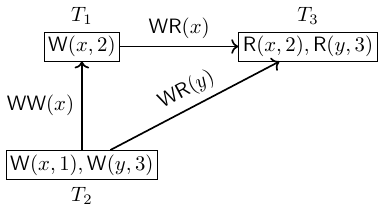}
				\captionsetup{skip=2pt}
		\caption{A serializable \uniqueval{} history.}
		\label{fig:ex-ser-uv-depgraph}
	\end{subfigure}
	\hfill
	\begin{subfigure}[b]{0.5\columnwidth}
		\centering
		\includegraphics[width=\textwidth]{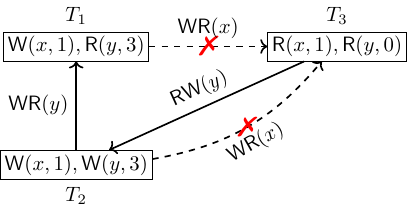}
				\captionsetup{skip=2pt}
		\caption{An unserializable \duplicateval{} history.}
		\label{fig:ex-nonser-dup-depgraph}
	\end{subfigure}
		\captionsetup{skip=4pt}
	\caption{Capturing \ser via dependency graphs.
		Solid arrows denote known dependencies, whereas dashed arrows indicate uncertain ones.}
	\label{fig:ex-ser-depgraph}
	\vspace{-2ex}
\end{figure}

\begin{example}
  \label{ex:ser-depgraph}
  Figure~\ref{fig:ex-ser-depgraph} illustrates the dependency-graph-based characterization of \ser using two example histories.  For simplicity, we omit $T_{\bot}$ in both cases.
	Figure~\ref{fig:ex-ser-uv-depgraph} shows (the existence of) a serializable dependency graph with edges  like 
	$T_2 \rel{\WW(x)} T_1$, 
	which are
	constructed from a \uniqueval{} history.

	In contrast, it is impossible to build an acyclic dependency graph from the \duplicateval{} history in Figure~\ref{fig:ex-nonser-dup-depgraph}.
	Specifically,
	for key $y$, we obtain the dependencies $T_2 \rel{\WR(y)} T_1$ and $T_3 \rel{\RW(y)} T_2$ (due to $T_{\bot} \rel{\WR(y)} T_3$ and $T_{\bot}  \rel{\WW(y)} T_2$).
Now	consider the transaction from which $\readevent(x, 1)$ in $T_3$ reads its value.
		This leads to two possible $\WR$ edges (shown as dashed arrows): $T_1 \rel{\WR(x)} T_3$ and $T_2 \rel{\WR(x)} T_3$.
However, in both cases, the resulting dependency graph contains a cycle.
\end{example}

\begin{figure*}[t]
	\centering
	\begin{subfigure}[b]{0.18\textwidth}
		\centering
		\includegraphics[width=\textwidth]{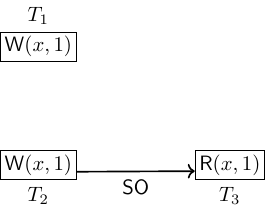}
		\captionsetup{skip=2pt}
		\caption{General history $\H$.}
		\label{fig:history-h2}
	\end{subfigure}
	\hfill
	\begin{subfigure}[b]{0.24\textwidth}
		\centering
		\includegraphics[width = 0.85\textwidth]{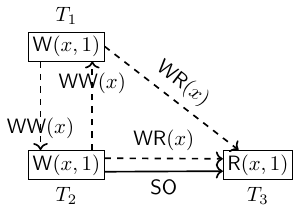}
		\captionsetup{skip=2pt}
		\caption{Hyper-polygraph $\G$ of $\H$.}
		\label{fig:rw-checking-h2-constraints}
	\end{subfigure}
	\hfill
	\begin{subfigure}[b]{0.28\textwidth}
		\centering
		\includegraphics[width = 0.70\textwidth]{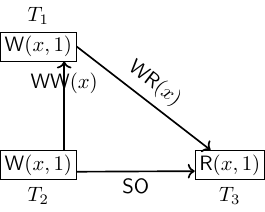}
		\captionsetup{skip=2pt}
		\caption{An acyclic graph compatible with $\G$.}
		\label{fig:rw-checking-h2-solution}
	\end{subfigure}
	\hfill
	\begin{subfigure}[b]{0.26\textwidth}
		\centering
		\includegraphics[width = 0.8\textwidth]{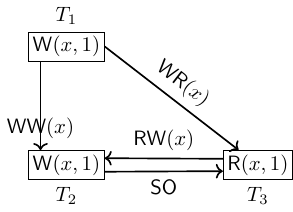}
		\captionsetup{skip=2pt}
		\caption{A cyclic graph compatible with $\G$.}
		\label{fig:rw-checking-h2-conflict}
	\end{subfigure}
	\captionsetup{skip=6pt}
	\caption{A general history $\H$, along with its hyper-polygraph $\G$
		and two compatible graphs.}
	\label{fig:rw-checking}
\end{figure*}

\inlsec{Polygraphs}
A dependency graph extending a history represents \emph{one} possible interpretation of the dependencies among the involved transactions, which may or may not satisfy \ser.
To capture \emph{all} possible dependency scenarios---allowing a verifier to check whether any of them satisfies \ser---state-of-the-art tools~\cite{Cobra:OSDI2020,PolySI:VLDB2023,Viper:EuroSys2023} utilize \emph{polygraphs}~\cite{SER:JACM1979,TIS:Book2002}.
Intuitively, a polygraph can be seen as a compact representation of a family of dependency graphs, each corresponding to a different possible internal execution consistent with the observed history. 

\begin{definition} \label{def:polygraph}
	A polygraph $\G = (\V, \E, \C)$ associated with a history $\H = (\T, \SO)$
	is a directed labeled graph $(\V, \E)$ called the \emph{known graph},
	together with a set $\C$ of \emph{constraints}, such that
	\begin{itemize}[leftmargin=20pt]
		\item $\V$ corresponds to all the transactions in $\H$;
		\item $\E = \set{(T, S, \SO) \mid T \rel{\SO} S} \cup \set{(T, S, \WR) \mid T \rel{\WR} S}$,
		where $\SO$ and $\WR$, when used as the third component in a tuple, serve as edge labels (i.e., types of dependencies); and
		\item $\C = \set{(T_{k}, T_{i}, \WW), (T_{j}, T_{k}, \RW)
			\mid (T_{i} \rel{\WR(\keyxvar)} T_{j})
			\land T_{k} \in \WriteTx_{\keyxvar} \land T_{k} \neq T_{i} \land T_{k} \neq T_{j}}$.
	\end{itemize}
\end{definition}


\begin{example}
According to the above definition, the polygraph associated with the history in 	Figure~\ref{fig:ex-ser-uv-depgraph} consists of a known graph with  nodes $\set{T_1, T_2, T_3}$ and edges $\set{T_1 \rel{\WR} T_3, T_2  \rel{\WR} T_3}$, along with a set of two constraints  $\set{(T_{2}, T_{1}, \WW), (T_{3}, T_{2}, \RW)}$ capturing the uncertainty in version order between $T_1$ and $T_2$.

These constraints are then encoded as a SAT formula and solved using an SMT solver that supports  theories like graph acyclicity, as we will see in Section~\ref{section:baseline}.
The acyclic dependency graph that satisfy \ser in Figure~\ref{fig:ex-ser-uv-depgraph} can be viewed as a solution produced by this solving process.
\end{example}

%% file: sections/epg-nobi-0424.tex

\section{Hyper-Polygraphs}
\label{section:hpg}

A dependency graph represents one possible execution of database transactions.
A polygraph captures a family of such graphs by fixing the $\WR$ relation---an assumption that holds under \uniqueval{},  where each written value is unique and thus unambiguously matched by reads---while allowing uncertainty in the $\WW$ relation.
However, this assumption, relied upon by all existing verifiers, does not always hold in practice, particularly under general workloads where the same value may be written multiple times, making read-from mappings ambiguous.

To address this representation problem, we introduce \emph{hyper-polygraphs} as a generalization of polygraphs that additionally account for uncertainty in $\WR$. Intuitively, a hyper-polygraph is a family of polygraphs, each corresponding to a distinct resolution of the $\WR$ relations, i.e., one polygraph per read-from mapping.




\begin{definition} \label{def:hyper-polygraph}
  A hyper-polygraph $\G = (\V, \E, \C)$
  for a history $\H = (\T, \SO)$
  is a directed labeled graph $(\V, \E)$, referred to as the known graph,
  together with \emph{a pair of constraint sets} $\C = (\wwcons, \wrcons)$, where
  \begin{itemize}[leftmargin=15pt]
    \item $\V$ is the set of nodes, corresponding to the transactions in $\H$;
    \item $\E \subseteq \V \times \V \times \type \times \Key$
      is a set of edges,
      where each edge is labeled with a dependency type from
      $\type = \set{\SO, \WR, \WW, \RW}$ and a key from $\Key$;\footnote{
For edges of type \(\SO\), the key component is irrelevant.
      }
    \item $\wwcons$ is a constraint set over  uncertain version orders, defined as
$\wwcons = \Bset{\btuple{T \rel{\WW(x)} S, S \rel{\WW(x)} T}
      \mid T \in \WriteTx_{x} \land S \in \WriteTx_{x} \land T \neq S}$; and
    \item $\wrcons$ is a constraint set over  uncertain read-from mappings, defined as
 $\wrcons = \Bset{\bigcup\limits_{T_{i} \vdash \writeevent(x, v)} \set{T_{i} \rel{\WR(x)} S}
      \mid S \vdash \readevent(x, v)}$.
  \end{itemize}
\end{definition}

Given two transactions $T$ and $S$, a type $\typevar \in \type$,
and a key $x \in \Key$,
we also write the edge $(T, S, \typevar, x)$ as $T \rel{\typevar(x)} S$.

\begin{example}
  \label{ex:hyper-polygraph}
  Consider the general history $\H$ shown in Figure~\ref{fig:history-h2}.
  It consists of three transactions $T_{1}$, $T_{2}$, and $T_{3}$, with a session order edge $T_2 \rel{\SO} T_3$.
  Both $T_{1}$ and $T_{2}$ write the same value $1$ to the same key $x$.
   The hyper-polygraph $\G$ constructed from $\H$ includes the following two constraint sets:
  \begin{itemize}
    \item $\wwcons = \Bset{\btuple{T_1 \rel{\WW(x)} T_2, T_2 \rel{\WW(x)} T_1}}$,
    representing the uncertainty in version order between $T_1$ and $T_2$ (shown as dashed edges in Figure~\ref{fig:rw-checking-h2-constraints});
    \item $\wrcons = \Bset{(T_1 \rel{\WR(x)} T_3,\ T_2 \rel{\WR(x)} T_3)}$, 
    representing the uncertainty in read-from mappings for $T_3$ (also shown as  dashed edges in Figure~\ref{fig:rw-checking-h2-constraints}).
  \end{itemize}
\end{example}


\inlsec{Characterizing Serializability using Hyper-Polygraphs}
A hyper-polygraph for a \duplicateval{} history can be viewed as a family of dependency graphs that are \emph{compatible with} it, where each graph resolves the constraints by selecting exactly one $\WW$ or $\WR$ edge from the $\wwcons$ and $\wrcons$ constraint sets, respectively.

\begin{definition} \label{def:compatible-graphs-with-a-hyper-polygraph}
  A directed labeled graph $\G' = (\V', \E')$ is compatible with
  a hyper-polygraph $\G = (\V, \E, \C)$ if
  \begin{itemize}[leftmargin=25pt]
    \item $\V' = \V$;
    \item $\E' \supseteq \E$ such that
      $\forall x \in \Key.\; \forall T, T', S \in \V'.\;
       (T', T, \WR, x) \in \E' \land
        (T', S, \WW, x) \in \E' \implies (T, S, \RW, x) \in \E'$;
    \item $\forall \consvar \in \wwcons.\;
      |\E' \cap \consvar| = 1$; and
    \item $\forall \consvar \in \wrcons.\;
      |\E' \cap \consvar| = 1$.
  \end{itemize}
\end{definition}

\begin{example}
  \label{ex:compatible-graphs}
  Figures~\ref{fig:rw-checking-h2-solution} and
  \ref{fig:rw-checking-h2-conflict} depict two graphs
  compatible with the hyper-polygraph $\G$ of the history $\H$.
  In Figure~\ref{fig:rw-checking-h2-solution},
  $T_{3}$ reads $x$ from $T_{1}$, which overwrites the value written by $T_{2}$.
  In contrast,
in Figure~\ref{fig:rw-checking-h2-conflict},
  $T_{3}$ reads $x$ from $T_{1}$, but the value is overwritten by $T_{2}$, resulting in  $T_{3} \rel{\RW(x)} T_{2}$.
\end{example}

Based on Theorem~\ref{thm:ser-depgraph},
we  obtain the following hyper-polygraph-based characterization of \ser;
the proof is provided in Appendix~\ref{ss:appendix-proof-hpg-ser}.

\begin{theorem}
  \label{thm:ser-hpg}
  A history $\H$ satisfies \emph{\ser{}} if and only if
  \emph{$\H \models \intaxiom$} and
  there exists an acyclic graph compatible with
  the hyper-polygraph of $\H$.
\end{theorem}



\begin{example}
  \label{ex:ser-characterization}
The graph compatible with $\G$ in Figure~\ref{fig:rw-checking-h2-conflict} contains a cycle: $T_3 \rel{\RW(x)} T_2 \rel{\SO} T_3$.
In contrast, the compatible graph in Figure~\ref{fig:rw-checking-h2-solution} is acyclic.
According to Theorem~\ref{thm:ser-hpg}, we conclude that $\H_3$ satisfies \ser.
\end{example}


%% file: sections/alg-nobi-0424.tex





\input{sections/ser-smt-baseline}

\input{sections/ser-smt-hypo}

\input{sections/ser-smt-dedicated}

%% file: sections/ser-smt-baseline.tex

\section{Off-the-Shelf SMT Solving: A Baseline}
\label{section:baseline}


To begin, we present a strong baseline approach for verifying \ser using MonoSAT~\cite{MonoSAT:AAAI2015}, an off-the-shelf SMT solver optimized for checking graph properties such as acyclicity.
MonoSAT serves as the core engine for  all state-of-the-art SMT-based isolation verifiers, including Cobra, PolySI, and Viper. 

Given a history $\H$ and following the workflow in Figure~\ref{fig:black-box-arch}, we first construct its hyper-polygraph. This involves extracting the transaction set $\T$, the session order $\SO$, and unique $\WR$ dependencies for the known graph, along with possible $\WW$ and $\WR$ dependencies as constraints.
We then focus on the two key steps: \emph{encoding} the hyper-polygraph into SAT formulas (Section~\ref{ss:encoding}) and \emph{solving} them with MonoSAT (Section~\ref{ss:solving}). 
We illustrate each step using the example history $\H$ shown in Figure~\ref{fig:history-h2}.


\paragraph{Preliminaries}
In propositional logic, a boolean variable $v$ can take the value $\true$ or $\false$.
A \emph{literal} $l$ refers to a variable $v$ or its negation $\lnot v$.
A \emph{clause} $C$ is a disjunction of one or more literals,
e.g., $C = l_1 \lor l_2 \lor \cdots \lor l_n$.
A \emph{formula} $\F$ is constructed using literals and the logical connectives $\land$, $\lor$, $\lnot$, or $\implies$.
Typically, we represent formulas in conjunctive normal form (CNF),
where $\F$ is a conjunction of multiple clauses:
$\F = C_1 \land C_2 \land \cdots  \land C_m$.
An \emph{assignment} of a variable $v$ is either $\true$ or $\false$, represented as 
$v$ or $\lnot v$, respectively.
A \emph{model} $\M$ for a formula $\F$ is a set of assignments to the
	boolean vairables appearing in $\F$ that makes  $\F$ $\true$.
	If $\M$ does not specify assignments for all variables in $\F$, 
	 it is referred to as a \emph{partial model}.



\subsection{Encoding}
\label{ss:encoding}

We represent the existence of each edge in the hyper-polygraph $\G = (\V, \E, \C)$ using a boolean variable. 
First, the known graph of $\G$ is encoded by the formula
\[
\small 
\Phi_{\mathit{KG}} = \csovar_{2, 3}
\]
	where the variable $\csovar_{2, 3}$, set to  $\true$, 
	indicates the presence of the $\SO$ edge $T_2 \rel{\SO} T_3$.

Second, the $\wwcons$ constraints of $\G$ are encoded by the formula
\[
	\small
\begin{aligned}
	\Phi_{\wwcons} = \;&(\cwwvar_{1,2}^x \lor \cwwvar_{2,1}^x)\land
	(\lnot \cwwvar_{1,2}^x \lor \lnot \cwwvar_{2,1}^x)
\end{aligned}
\]
where the variables $\cwwvar_{1,2}^x$ and $\cwwvar_{2,1}^x$ represent the existence of the edges
$T_1 \rel{\WW(x)} T_2$ and $T_2 \rel{\WW(x)} T_1$, respectively.
This formula enforces that exactly one of the two variables is assigned $\true$, ensuring a total order between the two conflicting writes.


Moreover, the $\wrcons$ constraints of $\G$ are encoded by the formula
\[
\small
\begin{aligned}
    \Phi_{\wrcons} = \; &(\cwrvar{x}_{1,3} \lor \cwrvar{x}_{2,3}) \land
(\lnot \cwrvar{x}_{1,3} \lor \lnot \cwrvar{x}_{2,3})
\end{aligned}
\]
where the variables $\cwrvar{x}_{1,3}$ and $\cwrvar{x}_{2,3}$ denote the presence of the edges
$T_1 \rel{\WR(x)} T_3$ and $T_2 \rel{\WR(x)} T_3$, respectively.
This formula enforces that exactly one of these edges holds, ensuring a unique read-from relation for $T_3$.

Finally, the encoding of the derivation rule for $\RW$ edges in $\G$ is captured by the formula
\[
\small
\begin{aligned}
	\Phi_{\RW} = \; & (\cwwvar_{1,2}^x \land \cwrvar{x}_{1,3} \implies \;\crwvar_{3,2}^x) \;\land  (\cwwvar_{2,1}^x \land \cwrvar{x}_{2,3}\implies\;\crwvar_{3,1}^x) \\
	\Leftrightarrow\; 
	& (\lnot \cwwvar_{1,2}^x \lor \lnot \cwrvar{x}_{1,3} \lor \crwvar_{3,2}^x) \;\land (\lnot \cwwvar_{2,1}^x \lor \lnot \cwrvar{x}_{2,3} \lor \crwvar_{3,1}^x)
\end{aligned}
\]
where $\crwvar_{3,1}^x$ and $\crwvar_{3,2}^x$ represent the existence of the $\RW(x)$ edges $T_3 \rel{\RW(x)} T_1$ and $T_3 \rel{\RW(x)} T_2$, respectively. 
This encoding ensures that an $\RW$ edge is derived only when both the corresponding $\WW$ and $\WR$ edges are present.

Overall, the complete encoding for $\G$ is given by
\[
\small 
\Phi_{\G} = \Phi_{\mathit{KG}} \land \Phi_{\wwcons} \land \Phi_{\wrcons} \land \Phi_{\RW}.
\]

\subsection{Solving}
\label{ss:solving}

Each satisfying assignment of $\Phi_{\G}$ corresponds to a graph $\G'$ that is compatible with  $\G$.
To ensure that $\G'$ is acyclic, we assert the predicate $\acyclicencoding(\G')$ in MonoSAT.
Hence, the history $\H$ is serializable if and only if the formula 
$\;\Phi_{\G} \land \acyclicencoding(\G')\;$ is satisfiable.




MonoSAT implements the \emph{Conflict-Driven Clause Learning with Theory} (CDCL(T)) framework~\cite{SMT:JACM2006} to decide the satisfiability of SAT formulas augmented with theory predicates.
 It combines a \emph{SAT solver}, which searches for a satisfying assignment to the boolean formula (e.g., $\Phi_{\G}$), with a \emph{theory solver}, which ensures that the asserted predicates (e.g., $\acyclicencoding(\G')$) hold under the current assignment.
 As the SAT solver assigns variables, the theory solver incrementally updates a graph $\ug$ by adding edges for variables set to $\true$. If a predicate is violated, a conflict clause is generated and learned by the SAT solver to prevent revisiting the same conflict.

\begin{figure}
  \centering
	\includegraphics[width=\columnwidth]{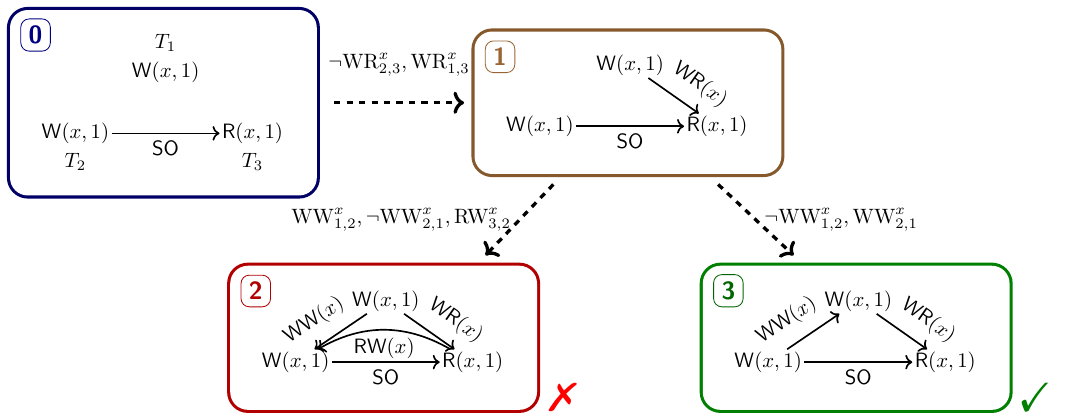}
	        		\captionsetup{skip=3pt}
	\caption{An example solving process for  $\H$ in Figure~\ref{fig:rw-checking}.}
	\label{fig:running-example-solving}
	\vspace{-2ex}
\end{figure}

The solving process may vary depending on the order in which variables are assigned.
Figure~\ref{fig:running-example-solving} illustrates one such procedure for $\G$, with the graph $\ug$ progressing through four stages.\footnote{We also illustrate in Appendix~\ref{section:appendix-cdclt} how the SAT solver and the theory solver collaborate in CDCL(T) for this example.}
Initially, only $\csovar_{2, 3}$ is assigned $\true$, yielding  \statebox[sblue]{0}. 
Suppose the SAT solver next assigns $\lnot \cwrvar{x}_{2, 3}$.
Then $\cwrvar{x}_{1,3}$ must be $\true$ to satisfy the clause  $\cwrvar{x}_{1,3} \lor \cwrvar{x}_{2,3}$.
This is also known as \emph{unit propagation}~\cite{DPLL:CACM1961} that occurs when a clause becomes unit, i.e.,
only one literal remains unassigned ($\cwrvar{x}_{1,3}$ in this case), forcing the literal to be assigned  $\true$ to satisfy the clause.
The theory solver adds the edge $T_1 \rel{\WR(x)} T_3$ to $\ug$, leading to \statebox[sbrown]{1}.
No cycles are detected at this point.

Next, assigning $\cwwvar_{1, 2}^x$ triggers $\lnot \cwwvar_{2, 1}^x$  due to  mutual exclusion,
and then $\crwvar_{3, 2}^x$ via the clause $\lnot \cwwvar_{1,2}^x \lor \lnot \cwrvar{x}_{1,3} \lor \crwvar_{3,2}^x$,
which
leads to the updated graph $\ug$ in \statebox[sred]{2}.
At this point,  the theory solver detects a cycle: $T_2 \rel{\SO} T_3 \rel{\RW(x)} T_2$ (see also Figure~\ref{fig:rw-checking-h2-conflict}).
It then returns a conflict clause $\lnot \crwvar_{3, 2}^x \lor \lnot \csovar_{2, 3}$ to the SAT solver, 
indicating that removing either edge would break the cycle.
The SAT solver then backtracks from $\cwwvar_{1, 2}^x$, assigns its negation $\lnot \cwwvar_{1, 2}^x$, and propagates $\cwwvar_{2, 1}^x$.
This results in 
 \statebox[sgreen]{3}, where all variables are assigned and no cycles are present.
Hence, an acyclic compatible graph is found (see also Figure~\ref{fig:rw-checking-h2-solution}), and the history $\H$ is serializable.

%% file: sections/ser-smt-hypo.tex
\section{Workload Characteristics for SMT Solving}
\label{section:characteristics-of-histories}


General-purpose SMT solvers like MonoSAT are agnostic to the characteristics of database  workloads, thereby missing opportunities for targeted optimization.
In contrast, we hypothesize that leveraging workload-specific characteristics can significantly improve solving efficiency for  isolation verification.
This section presents two key observations that motivate our tailored SMT strategies.

\subsection{Prevalence of 2-width Cycles}
\label{ss:observation-small-cycles}

	A key bottleneck arises in the solving stage from ``bad'' guesses that  introduce cycles in the dependency graph.
	When this occurs, the solver must backtrack and explore alternative choices, which can be costly.
To mitigate this, prior work~\cite{Cobra:OSDI2020,PolySI:VLDB2023,Viper:EuroSys2023}  introduces \emph{pruning},  
a preprocessing technique
 that identifies and precludes infeasible $\WW$ constraints by analyzing reachability \emph{before} the actual solving process starts.
For example, consider a $\WW$ constraint $\{T_1 \rel{\WW(x)} T_2, T_2 \rel{\WW(x)} T_1\}$. If $T_2$ is already reachable from $T_1$, then choosing $T_2 \rel{\WW(x)} T_1$ would result in a cycle. Hence, this choice can be safely pruned, eliminating the need for the solver to explore it further.
 In the context of SAT solving, pruning corresponds to adding a conflict clause with a single $\WW$ variable, e.g., $\lnot \cwwvar_{2, 1}^x$ in this case, to the encoding for $\Phi_{\G}$ (Section~\ref{ss:encoding}).

Essentially, pruning shifts the cost of resolving search-time conflicts to a pre-solving analysis phase.
This raises a natural question: can more of the solving overhead be offloaded to preprocessing?
In other words, is there a sweet spot that balances the cost of precomputation with the benefit of reduced solving effort, thereby improving overall performance?

To explore this, we adapt the notion of \emph{cycle width}~\cite{Zord:PLDI2021,Satisfiability:TOPLAS2023},
which we define as the number of \emph{unique} $\WW$ and $\WR$ variables appearing in a conflict clause.
While conventional pruning relies on unit clauses (i.e., 1-width cycles), broader pruning can be achieved by encoding larger cycles using multi-variable clauses.
To avoid redundancy, we count each variable only once and disregard $\SO$ variables, which are always $\true$, and $\RW$ variables, which can be derived from $\WW$ and $\WR$.

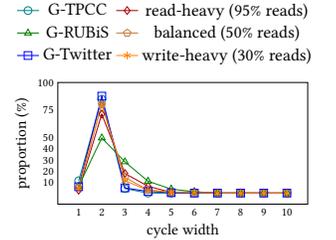
\begin{wrapfigure}{right}{4cm}
	\vspace{-3ex}
	\pgfplotsset{height=140pt, width=240pt}
	\pgfplotsset{tick style={draw=none}}
	\centering
		\begin{scaletikzpicturetowidth}{0.23\textwidth}
			\begin{tikzpicture}[scale=\tikzscale]
				\begin{axis}[
					xlabel={\LARGE cycle width},
					ylabel={\LARGE proportion (\%)},
					xlabel near ticks,
					ylabel near ticks,
					ymax=100,
					ytick={10,20,30,40,50,75,100},
					xtick={1,2,3,4,5,6,7,8,9,10,15,20},
					legend pos=north east,
					legend style={xshift=15pt, yshift=70pt,draw=none, legend columns=2, nodes={scale=1.5}}
					]
					\addplot[color=teal,mark=o,mark size=3pt] table [x=param, y=gtpcc, col sep=comma] {figs/cycle-width/data/minimized-data.csv};
					\addplot[color=red!65!black,mark=diamond,mark size=3pt] table [x=param, y=grh, col sep=comma] {figs/cycle-width/data/minimized-data.csv};
					\addplot[color=green!50!black,mark=triangle,mark size=3pt] table [x=param, y=grubis, col sep=comma] {figs/cycle-width/data/minimized-data.csv};
					\addplot[color=brown,mark=pentagon,mark size=3pt] table [x=param, y=grw, col sep=comma] {figs/cycle-width/data/minimized-data.csv};
					\addplot[color=blue,mark=square,mark size=3pt] table [x=param, y=gtwitter, col sep=comma] {figs/cycle-width/data/minimized-data.csv};
					\addplot[color=orange,mark=asterisk,mark size=3pt] table [x=param, y=gwh, col sep=comma] {figs/cycle-width/data/minimized-data.csv};
					
					\legend{G-TPCC, read-heavy (95\% reads), G-RUBiS, balanced (50\% reads), G-Twitter, write-heavy (30\% reads)}
				\end{axis}
			\end{tikzpicture}
		\end{scaletikzpicturetowidth}
	\captionsetup{skip=0pt}
	\caption{Cycle width distribution across conflicts.}
	\label{fig:cycle-width-minimal}
	\vspace{-2ex}
\end{wrapfigure}

We empirically measure the widths of conflict cycles encountered during solving across six representative benchmarks, including TPC-C and YCSB-like transaction  workloads with varying read proportions (see Section~\ref{ss:setup} for details).
Each benchmark consists of 1000 transactions.
For each conflict, we record the \emph{minimal} cycle width, as it
may correspond to multiple cycles depending on the search path. 
As shown in Figure~\ref{fig:cycle-width-minimal}, 2-width cycles dominate across all benchmarks. This suggests that many conflicts could be prevented by pre-encoding 2-variable clauses.
We leverage this insight in Section~\ref{ss:small-width-cycles}, incorporating 2-width cycle encodings into the SAT formula $\Phi_\G$ to strike a  balance between pre-solving effort and solving efficiency
(further validated in Section~\ref{sss:choosing-encode-width}).

\subsection{Polarity as Schedule Reconstruction}
\label{ss:observation-polarity}

Another major source of inefficiency in SMT solving stems from  poor polarity choices when assigning decision variables. 
A key insight we introduce is that 
each polarity decision---whether to include or exclude a dependency edge---implicitly defines a transaction schedule. 
Suboptimal choices can conflict with the actual, though hidden (due to the black-box setting), schedule enforced by the database,  leading to dependency cycles and costly backtracking.

General-purpose solvers such as MonoSAT treat all variables uniformly and assign polarity without regard to the execution context.
In contrast, we observe that many dependencies, including session order $\SO$ and uniquely determined read-from relation $\WR$, are already known from the observable history and can be leveraged to guide more informed polarity decisions.
 Ideally, if the internal transaction schedule were accessible, one could align polarity assignments perfectly to avoid conflicts. 
 While this is infeasible in black-box settings, the visible dependencies still capture a significant portion of the execution order. 
 We hypothesize that exploiting this partial order can substantially reduce conflict rates and backtracking overhead.
The following example illustrates this intuition.

\begin{figure}[h!]

	\centering
	\includegraphics[width=\columnwidth]{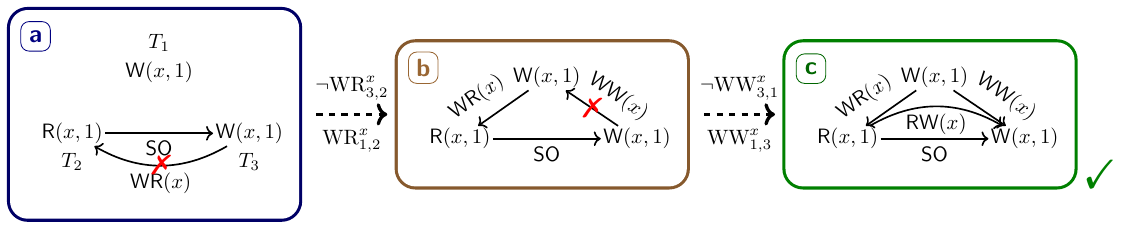}
	\captionsetup{skip=2pt}
	\caption{Guiding polarity picking via known dependencies.}
	\label{fig:example-polarity}
	\vspace{-2ex}
\end{figure}

\begin{example}
In Figure~\ref{fig:example-polarity}, the solver must resolve two constraints: $\{T_1 \rel{\WR(x)} T_2, T_3 \rel{\WR(x)} T_2\}$ and  $\{T_1 \rel{\WW(x)} T_3, T_3 \rel{\WW(x)} T_1\}$. 
Initially, in \statebox[sblue]{a}, the solver must decide between two possible read-from candidates for $T_2$.
By recognizing the known session order $T_2 \rel{\SO} T_3$, the solver correctly assigns $\cwrvar{x}_{3,2} = \false$, avoiding a potential cycle between $T_2$ and $T_3$.
The solver then proceeds to \statebox[sbrown]{b}, where it must resolve the $\WW$ constraint.
Again, guided by the inferred dependency $T_1 \rel{\WR(x)} T_3$, it avoids introducing a cycle, e.g., 
$T_1 \rel{\WR(x)} T_2 \rel{\SO} T_3 \rel{\WW(x)} T_1$,  by choosing $\cwwvar_{3,1}^x = \false$.

\end{example}

This example underscores the importance of polarity decisions: 
in large histories, a single incorrect polarity choice may introduce an edge that contradicts the actual schedule. 
Consequently, the solver is misled into exploring an exponentially larger search space, incurring substantial overhead due to cascading conflicts and backtracking.
Motivated by this insight, the next section introduces a polarity picking strategy---integrated into our dedicated SMT solver---that exploits the observable partial order in the execution history. 
We evaluate its impact in Section~\ref{subsubsec:impact-polarity}.

%% file: sections/ser-smt-dedicated.tex

\section{\ourtool with Tailored SMT Solving}\label{sec:alg}

\input{sections/ser-smt-overview}
\input{sections/opt-cycles}
\input{sections/opt-polarity}
\input{sections/complexity}





%% file: sections/ser-smt-overview.tex


\ourtool builds upon the baseline algorithm described in Section~\ref{section:baseline}.
Motivated by our observations in Section~\ref{section:characteristics-of-histories},
it incorporates two key optimizations: efficient handling of 2-width cycles (Section~\ref{ss:small-width-cycles}) and polarity picking for decision variables (Section~\ref{ss:polarity}).

\input{figs/algs/ser-smt-theory-framework}
\input{algs/theory-solver-main.tex}

Figure~\ref{fig:ser-smt-theory-framework} shows the high-level workflow of \ourtool, comprising four main stages: 
(i) \emph{constructing} the hyper-polygraph $\G$ from the input history $\H$, 
(ii) \emph{pruning} infeasible constraints in $\G$, 
(iii) \emph{encoding} the pruned graph $\G_p$ into a SAT formula $\formula$, 
and (iv) \emph{solving} $\formula$. 
The corresponding pseudo-code is given in Algorithm~\ref{alg:solver},
which we refer to alongside the figure to explain  the procedure.

The reminder of 
Algorithm~\ref{alg:solver} details the solving phase, specifically
 the interaction between the SAT frontend (\textsc{SatSolve}, Line~\ref{procedure:satsolve}) and the theory backend (\textsc{TheorySolve}, Line~\ref{procedure:theorysolve}). 
Let $\ug$ denote the current known graph maintained in the theory solver. 
Initially, $\ug$ is set to the  hyper-polygraph $\G$ constructed from $\H$. 
In each iteration, the SAT solver selects a decision variable $v$ {(Line~\ref{line:satsolve-call-choose-decision-var})} and its polarity, yielding a decision literal $l$ (Line~\ref{line:satsolve-call-pick-polarity}).
This literal is then unit propagated, producing a partial assignment $\pa$ (Line~\ref{line:satsolve-call-unit-propagate})
 that is passed to the theory solver for conflict checking  (Line~\ref{line:satsolve-call-theory-solve}).

Specifically,
the theory solver first {derives} $\RW$ edges from the current $\WW$ and $\WR$ edges {(Line~\ref{line:theorysolve-call-derive-rw-edges})} and checks whether the updated graph $\ug$ contains any cycles {(Line~\ref{line:theorysolve-call-cycle-detection})}.
If a cycle is found, it generates a conflict clause and returns it to the SAT solver (Line~\ref{line:theorysolve-return-conflict-clause}).
The SAT solver then resolves the conflict, either terminating with $\unsat$ (Line~\ref{line:satsolve-return-unsat}) or backtracking and updating $\pa$ (Line~\ref{line:satsolve-backtrack}).
If no cycle is detected, the solver continues by selecting the next decision variable.
Once all variables have been assigned without encountering conflicts (Line~\ref{line:satsolve-check-if-assignment-all}), the solver reports $\sat$ (or serializable). 

%% file: figs/algs/ser-smt-theory-framework.tex

\begin{figure}[h]
	\vspace{-2ex}
\includegraphics[width = .92\columnwidth]{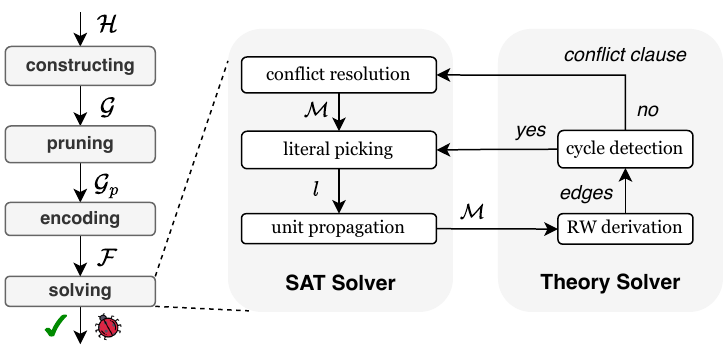}
		        		\captionsetup{skip=-5pt}
	\caption{Workflow of \ourtool.
	}
	\label{fig:ser-smt-theory-framework}
	\vspace{-2ex}
\end{figure}

%% file: algs/theory-solver-main.tex

\begin{algorithm}[t]
 \small
  \caption{The \ourtool algorithm}
  \label{alg:solver}
  \begin{algorithmic}[1]
  	\Statex{$\H$: the input history}
    \Statex{$\ug = (\V, \eug)$: 
    	 the current graph maintained in the theory solver; }
    \Statex{\quad initialized as the hyper-polygraph $\G$ constructed from $\H$ }
    \Statex{$\cvars$: a set of boolean variables managed by the SAT solver; initially $\emptyset$} 
    \Statex{$\cclauses$: a set of clauses managed by the SAT solver; initially $\emptyset$}
    \Statex{Both $\cvars$ and $\cclauses$ are constructed in \textsc{Encode} (Line~\ref{line:checkser-call-encode}).}
    \Statex{}

    \Function{VerifySER}{$\H$} 
      \label{procedure:checkser}
      \State $\G \gets \Call{Construct}{\H}$
        \label{line:checkser-call-construct}
        \Comment{\small  Alg.~\ref{algo:construct} in Appendix~\ref{section:appendix-alg}}
      \If{$\lnot \Call{Prune}{\G}$}
        \label{line:checkser-call-prune}
        \Comment{\small  Alg.~\ref{algo:fast-prune} in Appendix~\ref{section:appendix-alg}; see also Section~\ref{sss:prune-1-width-cycles}}
        \State \Return \false
      \EndIf
        \State $\formula \gets \Call{Encode}{\G_p}$
          \label{line:checkser-call-encode}
          \Comment{\small  Alg.~\ref{algo:encode} in Appendix~\ref{section:appendix-alg}; see also  Section~\ref{sss:encode-2-width-cycles}}
        \State \Return $\Call{Solve}{\formula, \G_p}$
          \label{line:checkser-call-solve}
    \EndFunction

    \Statex

    \Function{Solve}{$\formula, \G$}
      \label{procedure:solve}
      \ForAll{$T \rel{\typevar(x)} S \in \E$}
        \State{$\eug \gets \eug \cup \set{T \rel{} S}$} \label{line:solver-init-add-known-graph}
      \EndFor
      \label{line:solve-init-theory-solver}
      \State{\Return \Call{SatSolve}{$\formula, \G$}}
    \EndFunction
    
    \Statex
    \Function{SatSolve}{$\formula, \G$}
      \label{procedure:satsolve}
      \State{$\pa_{\mathit{total}} \gets \emptyset$}
      \While{$\cardinality{\pa_{\mathit{total}}} \neq \cardinality{\cvars}$}
      \label{line:satsolve-check-if-assignment-all}
        \State{$v \gets \Call{ChooseDecisionVar}{\pa_{\mathit{total}}, \cvars}$}
          \label{line:satsolve-call-choose-decision-var}
        \State{$\clit \gets \Call{PickPolarity}{v}$}
          \label{line:satsolve-call-pick-polarity}
          \Comment{\small Alg.~\ref{alg:pick-polarity} in Section~\ref{ss:polarity}}
          \State{$\pa \gets \Call{UnitPropagate}{\clit, \formula}$}
            \label{line:satsolve-call-unit-propagate}
          \State{$(\mathit{ret}, \cconflictcl) \gets \Call{TheorySolve}{\pa}$}
            \label{line:satsolve-call-theory-solve}
            \If{$\lnot \mathit{ret}$} \Comment{a conflict is detected}
            \If{$\lnot \Call{ResolveConflict}{\pa_{\mathit{total}}, \pa, \cconflictcl}$}
              \State{\Return $\unsat$}
              \label{line:satsolve-return-unsat}
              \Else
              \State{$\pa_{\mathit{total}} \gets \Call{Backtrack}{\pa_{\mathit{total}}, \pa, \cconflictcl}$}
              \label{line:satsolve-backtrack}
              \State{\textbf{continue}, \textbf{goto} Line~\ref{line:satsolve-check-if-assignment-all}}
              \label{line:continue}
            \EndIf
          \EndIf

        \State{$\pa_{\mathit{total}} \gets \pa_{\mathit{total}} \cup \pa$}
      \EndWhile
      \State{\Return $\sat$}
      \label{line:satsolve-return-sat}
    \EndFunction

    \Statex
    \Function{TheorySolve}{$\pa$}
      \label{procedure:theorysolve}
      \State{$\cedges \gets $\Call{DeriveRWEdges}{$\pa$}} 
          \label{line:theorysolve-call-derive-rw-edges}
      \If{$(\eug \cup \cedges)$ is cyclic} \Comment{\small use the PK algorithm~\cite{PKTopo:ACMJEA2006}}
        \label{line:theorysolve-call-cycle-detection}
        \State{$\cconflictcl \gets $ \Call{GenConflictClause}{$\eug, \cedges$}}
        \label{line:theorysolve-generate-conflict-clause}
        \State{\Return $(\false, \cconflictcl)$}
        \label{line:theorysolve-return-conflict-clause}
      \Else
        \State{$\eug \gets \eug \cup \cedges$}
        \State{\Return $(\true, \mathit{null})$}
      \EndIf
    \EndFunction
  \end{algorithmic}
\end{algorithm}

%% file: sections/opt-cycles.tex
\subsection{Small-Width Cycle Preprocessing}
\label{ss:small-width-cycles}

This optimization 
aims to 
shift the burden of handling conflicts within the solver 
 to a lightweight pre-solving analysis that focuses on small-width cycles. 
Intuitively, this preprocessing step acts as a
``sanity check'' before solving: 
it rules out choices that are guaranteed
  to fail (i.e., those that create cycles as exemplified in Section~\ref{ss:observation-small-cycles}), preventing the solver from wasting effort on them later.
It consists of two components:
(i) an aggressive pruning phase that eliminates all 1-width cycles through reachability analysis, and
(ii) a proactive encoding of 2-width cycles into the SAT formula, enabling early conflict detection via unit propagation during solving.







\subsubsection{Eliminating 1-Width Cycles}
\label{sss:prune-1-width-cycles}

We extend the pruning techniques from prior work~\cite{Cobra:OSDI2020,Viper:EuroSys2023,PolySI:VLDB2023},  which focus on $\WW$ constraints, 
by also pruning $\WR$ constraints and their derived $\RW$ counterparts.
The following examples illustrate the core ideas of our approach.

\paragraph{Case I: Pruning $\WW$ Constraints}
Consider the two transactions $T$ and $S$ in Figure~\ref{fig:pruning-ww-case},  both writing to the same key. If $T$ is already reachable from $S$ in the current known graph, then adding $T \rel{\WW(x)} S$ would introduce a 1-width cycle. This infeasible option can thus be pruned; the valid edge $S \rel{\WW(x)} T$ is added to the known graph.

\paragraph{Case II: Pruning $\WR$ Constraints}
We can similarly prune a $\WR$ edge if its inclusion would introduce a 1-width cycle in the known graph.
Consider the scenario in Figure~\ref{fig:pruning-wr-case}, where a transaction $S$ reads from key $x$ and has $n$ candidate writers $T_1, \dots, T_n$.
If $n - 1$ of these candidates have been eliminated due to cycle formation,  the remaining writer must be the only feasible source.
In this case, we can safely add the edge $T_i \rel{\WR(x)} S$  to the known graph.

\paragraph{Case III: Pruning via Derived $\RW$ Edges}
Derived edges may also lead to cycles. In Figure~\ref{fig:pruning-ww-case-rw-conflict}, assume that adding $T \rel{\WW(x)} T'$ induces $S \rel{\RW(x)} T'$ (via $T \rel{\WR(x)} S$). If $S$ is already reachable from $T'$, this creates a cycle, so $T \rel{\WW(x)} T'$ is pruned.

\ourtool iteratively applies this pruning process until the known graph stabilizes,
i.e., all 1-width cycles have been eliminated. 
The complete pseudo-code is provided in Algorithm~\ref{algo:fast-prune}
of Appendix~\ref{section:appendix-alg}.

\input{figs/pruning.tex}

\subsubsection{Encoding 2-width Cycles}
\label{sss:encode-2-width-cycles}

To  prevent conflicts caused by 2-width cycles during solving, \ourtool  encodes them into  SAT formulas.
The corresponding pseudo-code is provided in Algorithm~\ref{algo:init-pair-conflict}. 
We consider two  scenarios, as also illustrated in Figure~\ref{fig:pair-conflict}.

\paragraph{Case I: Canonical 2-Width Cycles}
This case illustrates a canonical 2-width cycle formed by two known paths, as shown in Figure~\ref{fig:pair-conflict-common}.
Let $T \leadsto T'$ and $S' \leadsto S$ be known dependencies in the graph, where $ \leadsto $ denotes reachability between  transactions.
 Adding both candidate edges $S \rel{} T$ and $T' \rel{} S'$ would complete a cycle.
We prevent this by adding the clause $\lnot v_{S, T} \lor \lnot v_{T’, S’}$ to the formula, where $v_{S, T}$ and $v_{T’, S’}$ are the corresponding boolean variables.

\paragraph{Case II: Derived $\RW$ Cycles}
In Figure~\ref{fig:pair-conflict-rw}, assume $T' \leadsto S$ is known. Adding both $T \rel{\WR(x)} S$ and $T \rel{\WW(x)} T'$ would derive $S \rel{\RW(x)} T'$, forming a cycle. 
To avoid this, we encode the clause:
$\lnot \cwwvar_{T, T'}^x \lor \lnot \cwrvar{x}_{T, S}$.

While longer cycles could, in theory, also be encoded, 
enumerating clauses over $k$ boolean variables incurs a time complexity of $O(|\cvars|^{k})$, making it impractical for large $k$.
Our analysis shows that 2-width clauses capture the majority of conflicts effectively (Figure~\ref{fig:encode-cycle-width}).
Hence, we focus on encoding only 2-width cycles, striking a  balance between preprocessing overhead and solving efficiency.

\vspace{-2ex}
	\subsubsection{Why our preprocessing works}
	Our encoding of 2-width cycles does not introduce additional constraints, 
	as the encoded cycles cannot appear in a valid model. 
Hence, equisatisfiability with the original encoding is preserved. 
Moreover, pruning can be viewed as a specialization of encoding 1-width cycles.
	 This process is lifted from solving to the manipulation of hyper-polygraphs,
	 and 
	 involves
	  two key steps: 
	  (i) eliminating infeasible edges in $\C$ (a specialization of conflict detection and learning), and (ii) adding the final edge to the known graph $(\V, \E)$ (a specialization of propagation). 
Correctness of pruning also follows from equisatisfiability with the original encoding. 
We defer the formal proofs to Appendix~\ref{ss:appendix-correctness-cycle}.

\input{algs/init-pair-conflict.tex}
\input{figs/pair-conflict.tex}

%% file: figs/pruning.tex

\begin{figure}[t]
  \centering
  \begin{subfigure}[b]{0.13\textwidth}
    \centering
    \includegraphics[width=\textwidth]{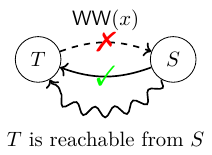}
     		        		\captionsetup{skip=4pt}
    \caption{Pruning $\WW$}
    \label{fig:pruning-ww-case}
  \end{subfigure}
  \hfill 
  \begin{subfigure}[b]{0.16\textwidth}
    \centering
    \includegraphics[width=\textwidth]{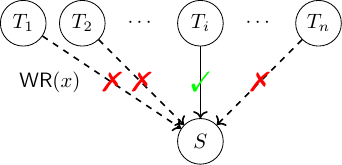}
     		        		\captionsetup{skip=4pt}
    \caption{Pruning $\WR$}
    \label{fig:pruning-wr-case}
  \end{subfigure}
  \hfill 
  \begin{subfigure}[b]{0.15\textwidth}
    \centering
    \includegraphics[width=\textwidth]{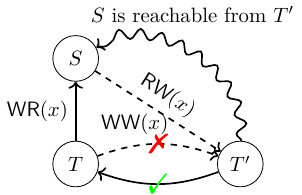}
     		        		\captionsetup{skip=4pt}
    \caption{Pruning via $\RW$}
    \label{fig:pruning-ww-case-rw-conflict}
  \end{subfigure}
  		        		\captionsetup{skip=2pt}
  \caption{Cases for pruning 1-width cycles. 
   }
  \label{fig:pruning-cases}
  \vspace{-2ex}
\end{figure}

%% file: algs/init-pair-conflict.tex
\begin{algorithm}[t]
\small
  \caption{Encoding 2-width cycles}
  \label{algo:init-pair-conflict}
  \begin{algorithmic}[1]
    \Function{EncodingCyclesOfWidth2}{$\V, \E$}
      \State{$\creachability \gets $ \Call{Reachability} {$\V, \E$}}
      \ForAll{$v_1, v_2 \neq v_{1} \in \cvars$}
        \If{\Call{CanonicalCycle}{$v_1, v_2, \creachability$} $\lor $ \Call{RWCycle}{$v_1, v_2, \creachability$}}
          \State{$\cclauses \gets \cclauses \cup \set{\lnot v_1 \lor \lnot v_2}$}
        \EndIf
      \EndFor
    \EndFunction
    \Function{CanonicalCycle}{$v_1, v_2, \creachability$}
      \State {\bf let} $(S, T) \gets v_1, (T', S') \gets v_2$
      \State{\Return $(T, \;T') \in \creachability \land (S, \;S') \in \creachability$}
    \EndFunction
    \Function{RWCycle}{$v_1, v_2, \creachability$}
      \State{\Return $\exists x.\;\exists T, S, T'.\;T' \neq T \land (T', S) \in \creachability\;\land\; 
              ((v_1 \triangleq \cwwvar_{T, T'}^x \land v_2 \triangleq \cwrvar{x}_{T, S}) \lor 
              (v_2 \triangleq \cwwvar_{T, T'}^x \land v_1 \triangleq \cwrvar{x}_{T, S}))$}
    \EndFunction
  \end{algorithmic}
\end{algorithm}

%% file: figs/pair-conflict.tex

\begin{figure}[t]
	  \vspace{-1ex}
  \centering
  \begin{subfigure}[b]{0.18\textwidth}
    \centering
    \includegraphics[width=\textwidth]{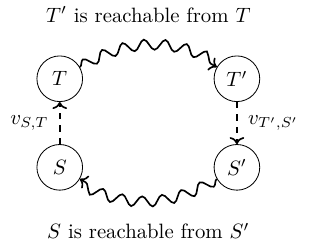}
         		        		\captionsetup{skip=4pt}
    \caption{Canonical  cycle}
    \label{fig:pair-conflict-common}
  \end{subfigure}
\hspace{6ex}
  \begin{subfigure}[b]{0.18\textwidth}
    \centering
    \includegraphics[width=\textwidth]{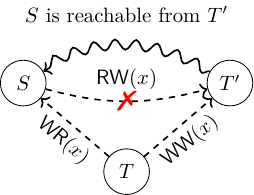}
         		        		\captionsetup{skip=4pt}
    \caption{Derived $\RW$ cycle}
    \label{fig:pair-conflict-rw}
  \end{subfigure}
       		        		\captionsetup{skip=2pt}
  \caption{Cases for encoding 2-width cycles.}
  \label{fig:pair-conflict}
  \vspace{-2ex}
\end{figure}

%% file: sections/opt-polarity.tex

\subsection{Order-Guided Polarity Picking}
\label{ss:polarity}

\ourtool leverages the current partial order of the dependency graph to guide polarity picking, as motivated in Section~\ref{ss:observation-polarity}. 
To achieve this,  we incorporate the \emph{pseudo-topological order} maintained by the PK algorithm~\cite{PKTopo:ACMJEA2006}, which is originally developed for dynamic cycle detection and also adopted by MonoSAT.
 The full polarity picking procedure is given in Algorithm~\ref{alg:pick-polarity}.

\input{algs/pick-polarity}

The PK algorithm dynamically maintains a level $\clevel{\cdot}$ for each vertex, which approximates its position in a topological ordering.
When a candidate edge $(\fromvar, \tovar)$ is considered, the algorithm first checks whether $\clevel{\fromvar} < \clevel{\tovar}$.
If so, the edge is guaranteed not to introduce a cycle and is immediately accepted.
If not, the algorithm performs a reachability check to ensure that no cycle would be formed, and if the edge is added, it updates the levels of affected vertices to maintain consistency.

This pseudo-topological order reflects the known partial order in the current  graph: every edge 
included satisfies $\clevel{\fromvar} < \clevel{\tovar}$.
 We exploit this order to guide polarity selection.
  For a decision variable $v$ representing a candidate edge $(\fromvar, \tovar)$,  we assign it a \emph{positive polarity} (i.e., include the edge) when $\clevel{\fromvar} < \clevel{\tovar}$, and a \emph{negative polarity} (i.e., exclude the edge) otherwise.

This heuristic encourages decisions that respect the observed partial order and are thus less likely to introduce cycles.
Although it cannot guarantee cycle-freedom in all cases, especially when the topological approximation is incomplete, it serves as a practical guide for reducing backtracking.
If a polarity choice later results in a conflict, the solver will backtrack and try the opposite polarity.
Since this heuristic does not alter the search space of CDCL(T), correctness is preserved.
The proof can be found in Appendix~\ref{ss:appendix-correctness-polarity}.

%% file: algs/pick-polarity.tex

\begin{algorithm}[h]
\small 
  \caption{Pseudo-topological-order-guided polarity picking}
  \label{alg:pick-polarity}
  \begin{algorithmic}[1]
    \Function{PickPolarity}{$v$}
      \State {\bf let} $(\fromvar, \tovar) \gets v$
      \If {$\clevel{\fromvar} < \clevel{\tovar}$}
        \State{\Return $v$}  \Comment{\small positive polarity}
      \Else
        \State{\Return $\lnot v$} \Comment{\small negative polarity}
      \EndIf
    \EndFunction
  \end{algorithmic}
\end{algorithm}

%% file: sections/complexity.tex
\subsection{Complexity Analysis} \label{ss:complexity}

\inlsec{Constructing}
Let $\totalsize{\C} = \sum\limits_{\ccons \in \wwcons \cup \wrcons} \cardinality{\ccons}$. 
The construction phase requires enumerating  all edges in $\C$, 
yielding a time complexity of $\bigo(\totalsize{\C})$. 
Let $k$ be the number of keys and $n$ the number of transactions.
In the worst case, $\bigo(\totalsize{\C}) = \bigo(kn^2)$. 

\inlsec{Pruning}
Pruning proceeds in iterative passes.
Each pass first computes reachability and then enumerates constraints in $\C$ to test the three pruning cases shown in Figure~\ref{fig:pruning-cases}. 
Computing reachability costs 
$\bigo\big(\cardinality{\V} \cdot \cardinality{\E}\big)$, 
where $(\V, \E)$ is the known graph. 
With reachability cached, testing each pruning case costs only $\bigo(1)$.
The main bottleneck is enumerating possible $\RW$ edges,
which can be  characterized by the number of  tuples 
$(T, T', S, x)$ with
$T, T', S \in \T, x \in \Key, T, T' \vdash \writeevent(x, \_)$ 
and $S \vdash \readevent(x, \_)$.
This costs
$\bigo(\sum\limits_{x \in \Key} \cardinality{\WriteTx_x}^2 \cdot \cardinality{\ReadTx_x})$, 
where $\ReadTx_x = \set{T \mid T \vdash \readevent(x, \_)}$. 
Let $N_{\P}$ be the number of pruning passes.
The overall complexity of pruning is then 
$\bigo\bigl(N_{\P} \cdot \big(\cardinality{\V} \cdot \cardinality{\E} + 
\sum\limits_{x \in \Key} \cardinality{\WriteTx_x}^2 \cdot \cardinality{\ReadTx_x}\big)\bigr)$.
Experiments show that $N_{\P}$ is typically small 
(less than $10$), and can thus be 
 treated as $\bigo(1)$.
In the worst case, pruning has complexity $\bigo(kn^3)$.

\inlsec{Encoding}
Let $\C_p$ denote the  constraints in the pruned hyper-polygraph $\G_p$.
The encoded formula consists of two parts: 
the constraints in $\C_p$ and the clauses for 2-width cycles.
The bottleneck lies in enumerating edge pairs to detect 2-width cycles 
(see Algorithm~\ref{algo:init-pair-conflict}), 
which costs $\bigo(\totalsize{\C_p}^2)$.
Hence, the total  encoding complexity is $\bigo(\totalsize{\C_p}^2)$, or 
 $\bigo(k^2n^4)$ in the worst case.

\inlsec{Solving}
The final stage, SMT solving, is NP-hard in general. 
Although prior research has analyzed SMT solvers and established lower bounds in certain settings~\cite{ProofComplexity:CAV2018},
 the precise complexity of our setting remains open~\cite{KnuthSatisfiability:book2015}. 

Despite this inherent theoretical hardness, experiments show that our approach achieves high efficiency on practical workloads, where worst-case scenarios rarely arise (Section~\ref{section:exp}).
This reflects the motivating insight of our work: 
database workload characteristics can be leveraged to build efficient and practical verifiers.


%% file: sections/eval-nobi-0424.tex
\section{Experiments}
\label{section:exp}

We have implemented our algorithm in a tool called \ourtool,
built on top of \minisat~\cite{MiniSAT:SAT2004},
a minimalistic and open-source SAT solver. 
We selected \minisat due to its simplicity and extensibility,
which allowed us to integrate our optimization strategies effectively.

To support the optimization of small-width cycles (Section~\ref{ss:small-width-cycles}), we compute reachability in the known 
graph using dynamic programming. 
Specifically, \ourtool maintains 
a 0-1 matrix of size $|\T| \times |\T|$, where each entry records whether one transaction is reachable from another. By traversing nodes in reverse topological order, 
it incrementally updates each node's reachable set: 
for an edge $(\fromvar, \tovar)$, it unions the reachability set of $\tovar$ into that of $\fromvar$. 
We implement the matrix  using bit vectors and bitwise operations, providing a lightweight alternative to GPU-based solutions~\cite{Cobra:OSDI2020}.
Overall, \ourtool has approximately 5k lines of C++ code.

We conduct an extensive evaluation of \ourtool,
focusing  on its checking efficiency,
while also assessing its bug-finding effectiveness.
Specifically, we aim to answer the following  questions:

\begin{itemize}[leftmargin=20pt]
	\item \textbf{Efficiency:}
		How efficiently does \ourtool perform
		across various general workloads (Section~\ref{subsubsec:vs-baseline})?
		Can it outperform existing tools (Section~\ref{subsubsec:vs-sota})?
		Does it scale to large histories (Section~\ref{subsubsec:scale})?
		How do its individual components contribute to the overall performance
		(Section~\ref{subsec:decompose})?
	\item \textbf{Effectiveness:}
	  How effective is \ourtool at uncovering isolation bugs
		in production database systems,
		particularly those related to \duplicateval{}
		(Section~\ref{subsec:bugs})?
\end{itemize}

\input{sections/eval-setup}

\input{sections/eval-performance}

\input{sections/eval-dissecting}

\input{sections/eval-effectiveness}

%% file: sections/eval-setup.tex

\subsection{Benchmarks and Experimental Setup} \label{ss:setup}

We evaluate \ourtool and competing verifiers
using two categories of benchmarks.
The first category comprises YCSB-like transactional workloads,
produced by a parametric workload generator built on top of PolySI~\cite{PolySI:VLDB2023}. 
This generator supports several configurable parameters,
with default values shown in parentheses:
the number of client sessions (20),
the number of transactions per session (100),
the number of read/write operations per transaction (20),
the overall read proportion (50\%),
the total number of keys (5k),
and the proportion of keys with duplicate write values (50\%).


We characterize duplicate write values using a Zipfian distribution,
where the parameter $\theta$ (0.5) controls the degree of duplication,
and $N$ (100) defines the size of  value space.
We then generate a set of representative workloads,
including RH (read-heavy, 95\% reads), WH (write-heavy, 70\% writes),
and BL (balanced, 50\% reads).
These workloads  align with \uniqueval{} ones
commonly used in prior work.
In addition, we include HD,
a variant of BL with a high degree of duplication ($\theta = 1.5$).
Each workload is executed on a PostgreSQL v15 instance
to produce  transactional histories,
each containing at least 10k transactions and 80k operations.

The second category comprises application-level benchmarks:
TPC-C~\cite{TPCC}, a standard OLTP benchmark
configured with 1 warehouse, 10 districts, 30k customers, and 5 transaction types;
RUBiS~\cite{CRUBis}, an auction-based system with 20k users and 200k items; and
TwitterClone~\cite{CTwitter}, a blogging application with Zipfian key access patterns.
We use a PostgreSQL instance to generate general transactional histories
from these applications, which we refer to as G-TPCC, G-RUBiS, and G-Twitter, respectively.
Notably, both RUBiS and TwitterClone naturally generate duplicate write values.
However, prior work~\cite{Cobra:OSDI2020,PolySI:VLDB2023,Viper:EuroSys2023}
disabled this behavior to enforce the \uniqueval{} assumption.
In contrast, we preserve these duplicate-value semantics
to more accurately reflect real-world behavior.

Unless otherwise stated,
all experiments are performed on a local machine
equipped with an Intel 13th Gen i5 CPU and 32GB RAM.

%% file: sections/eval-performance.tex

\subsection{Performance Evaluation}
\label{subsec:eval}

\subsubsection{Comparison with Baselines}
\label{subsubsec:vs-baseline}
Our first set of experiments compares \ourtool
with the baseline algorithm (Section~\ref{section:baseline}) across various general workloads.
We also include a stronger variant,  baseline+P,
which incorporates the pruning technique for \WW{} constraints
(Section~\ref{ss:observation-small-cycles}).

The experimental results are shown in Figure~\ref{fig:various-runtime},
omitting data points that exceed the 60s timeout.
\ourtool consistently outperforms both the baseline and its stronger variant.
Specifically, under increased concurrency,
such as more sessions (a), more transactions per session (b),
more operations per transaction (c), and smaller key space (d),
the two baselines suffer from exponential verification time,
while \ourtool incurs only moderate overhead.
Furthermore, in scenarios with increased uncertainty in \WW{} and \WR{} dependencies, such as write-heavy workloads (e) and higher proportions of duplicate keys (f),
\ourtool remains the fastest while maintaining fairly stable verification efficiency.

In addition, we measure the memory usage of all three tools
under the same settings as in Figure~\ref{fig:various-runtime}.
The trends are similar:
\ourtool uses less memory than both baselines across a range of workloads.
Memory plots are provided in Appendix~\ref{section:appendix-memory}.

\input{figs/eval/various/runtime.tex}
\input{figs/eval/uv/runtime/runtime.tex}
\input{figs/eval/scalability/scala.tex}

\subsubsection{Comparison with State-of-the-Art}
\label{subsubsec:vs-sota}

Our second set of experiments evaluate \ourtool
against four state-of-the-art verifiers:
Cobra~\cite{Cobra:OSDI2020} for \ser,
PolySI~\cite{PolySI:VLDB2023} and Viper~\cite{Viper:EuroSys2023} for \si,
and dbcop~\cite{Complexity:OOPSLA2019} for both.
Since Cobra supports GPU acceleration for pruning,
we also include its GPU-enabled versions in our evaluation.
Note that dbcop is included as a representative of non-SMT-based verifiers,
which rely on graph traversal algorithms, 
such as depth-first search,  to detect cycles.

For a fair comparison, we use the BlindW-RW dataset
adopted in prior work~\cite{Cobra:OSDI2020,Viper:EuroSys2023}.
This dataset exclusively consists of \uniqueval{} histories,
as all the above verifiers are restricted to such input.
Each history contains an even mix of read-only and write-only transactions,
with each transaction comprising eight operations.


As shown in Figure~\ref{fig:uv-runtime},
\ourtool outperforms all competitors in verifying both isolation levels.
In particular,
with only 16k transactions, it achieves up to a 4.4x speedup, even compared to Cobra with GPU acceleration.
Interestingly, even with  a stronger H800 GPU  (compared to the  V100  used in the original experiments~\cite{Cobra:OSDI2020}), \ourtool still maintains the performance gap. 
This suggests that the primary bottleneck in  verification lies not  in GPU capability but in the solving stage,
which is
consistent with the fact that  
GPUs are mainly used to 
accelerate reachability analysis before solving.
For \si, \ourtool delivers up to a 4x speedup over PolySI
and substantially outperforms Viper and dbcop.
All these results highlight the benefits of our dedicated SMT solving.



\subsubsection{Scalability}
\label{subsubsec:scale}
Verifying large histories is highly desirable for increasing confidence in a database's fulfilment of promised isolation guarantee.
Following the previous experiments, we
evaluate \ourtool's scalability on large histories, each containing up to 100k transactions and 2 million operations with a total of  50k keys.\footnote{We used a server with an AMD EPYC 7H12 64-Core processor and 1TB  of memory.	}


As shown in  Figure~\ref{fig:scalability},
verifying  large histories is manageable for \ourtool
on modern hardware, e.g., \ser verification completes in under 5 hours using less than 64 GB of memory.
In addition, two expected trends are observed.
First, both runtime and memory usage increase  as the proportion of read operations decreases (from RH to BL to WH workloads).
This trend is primarily due to the increased uncertainty in $\WW$ dependencies, which expands the set of potential read-from candidates and ultimately amplifies the overall uncertainty.

Second, verifying \si incurs  higher overhead than \ser.
This is expected: while \ser requires only an acyclic dependency graph, \si permits certain cycles, making the analysis inherently more complex.
In practice, \ourtool   maintains an additional induced graph for \si, which is more expensive to construct than deriving $\RW$ edges alone as for \ser.
This gap is also evident in prior evaluations on \uniqueval{} histories:
Cobra verifies a history with 10k transactions against \ser in 15s~\cite{Cobra:OSDI2020},
whereas Viper requires over 400s to verify \si for the same amount of transactions~\cite{Viper:EuroSys2023}.
This  gap becomes even more pronounced in the presence of duplicate values.

%% file: figs/eval/various/runtime.tex
\pgfplotsset{height=135pt, width=200pt}
\pgfplotsset{tick style={draw=none}}
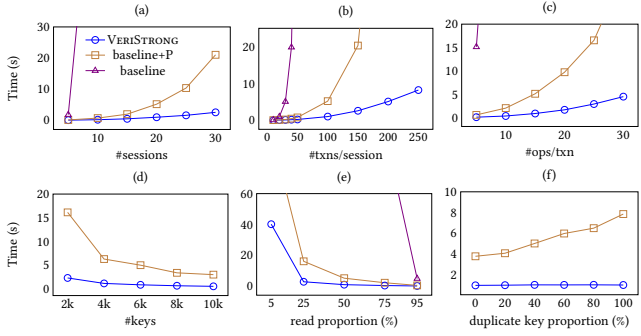
\begin{figure}[t]
	\centering
	\resizebox{\linewidth}{!}{
	\begin{scaletikzpicturetowidth}{0.315\textwidth}
		\begin{tikzpicture}[scale=\tikzscale]
			\begin{axis}[
				font=\LARGE,
				title={(a)},
				xlabel={\#sessions},
				ylabel={Time (s)},
				ymax=30,
				cycle multiindex* list={
								color       \nextlist
								mark list*  \nextlist
						},
				legend pos=north west,
				legend style={xshift=20pt,draw=none},
				]
				\addplot[color=blue,mark=o,mark size=3pt] table [x=param, y=ours, col sep=comma] {figs/eval/various/data/runtime-sessions.csv};
				\addplot[color=brown,mark=square,mark size=3pt] table [x=param, y=baseline-pruning, col sep=comma] {figs/eval/various/data/runtime-sessions.csv};
				\addplot[color=violet,mark=triangle,mark size=3pt] table [x=param, y=baseline, col sep=comma] {figs/eval/various/data/runtime-sessions.csv};
				\legend{\ourtool, baseline+P, baseline}
			\end{axis}
		\end{tikzpicture}
		\hspace{1ex}
	\end{scaletikzpicturetowidth}
	
	\begin{scaletikzpicturetowidth}{0.27\textwidth}
		\begin{tikzpicture}[scale=\tikzscale]
			\begin{axis}[
				font=\LARGE,
				title={(b)},
				xlabel={\#txns/session},
				ymax=25,
				legend pos=north west,
				]
				\addplot[color=blue,mark=o,mark size=3pt] table [x=param, y=ours, col sep=comma] {figs/eval/various/data/runtime-txns.csv};
				\addplot[color=brown,mark=square,mark size=3pt] table [x=param, y=baseline-pruning, col sep=comma] {figs/eval/various/data/runtime-txns.csv};
				\addplot[color=violet,mark=triangle,mark size=3pt] table [x=param, y=baseline, col sep=comma] {figs/eval/various/data/runtime-txns.csv};
			\end{axis}
		\end{tikzpicture}
				\hspace{1ex}
	\end{scaletikzpicturetowidth}
	
	\begin{scaletikzpicturetowidth}{0.27\textwidth}
		\begin{tikzpicture}[scale=\tikzscale]
			\begin{axis}[
				font=\LARGE,
				title={(c)},
				xlabel={\#ops/txn},
				ymax=20,
				legend pos=north west,
			]
      \addplot[color=blue,mark=o,mark size=3pt] table [x=param, y=ours, col sep=comma] {figs/eval/various/data/runtime-evts.csv};
      \addplot[color=brown,mark=square,mark size=3pt] table [x=param, y=baseline-pruning, col sep=comma] {figs/eval/various/data/runtime-evts.csv};
			\addplot[color=violet,mark=triangle,mark size=3pt] table [x=param, y=baseline, col sep=comma] {figs/eval/various/data/runtime-evts.csv};
  		\end{axis}
		\end{tikzpicture}
	\end{scaletikzpicturetowidth}
	}

	\resizebox{\linewidth}{!}{
			\begin{scaletikzpicturetowidth}{0.31\textwidth}
			\begin{tikzpicture}[scale=\tikzscale]
				\begin{axis}[
				font=\LARGE,
					title={(d)},
					xlabel={\#keys},
					ylabel={Time (s)},
					xtick={2000,4000,6000,8000,10000},
					xticklabels={2k,4k,6k,8k,10k},
					ymax=20,
					legend pos=north west,
					scaled ticks=false
					]
					\addplot[color=blue,mark=o,mark size=3pt] table [x=param, y=ours, col sep=comma] {figs/eval/various/data/runtime-keys.csv};
					\addplot[color=brown,mark=square,mark size=3pt] table [x=param, y=baseline-pruning, col sep=comma] {figs/eval/various/data/runtime-keys.csv};
					\addplot[color=violet,mark=triangle,mark size=3pt] table [x=param, y=baseline, col sep=comma] {figs/eval/various/data/runtime-keys.csv};
				\end{axis}
			\end{tikzpicture}
			\hspace{1ex}
		\end{scaletikzpicturetowidth}
		
	\begin{scaletikzpicturetowidth}{0.27\textwidth}
		\begin{tikzpicture}[scale=\tikzscale]
			\begin{axis}[
				font=\LARGE,
				title={(e)},
				xlabel={read proportion (\%)},
				xtick={5,25,50,75,95},
				ymax=60,
				legend pos=north west,
			]
      \addplot[color=blue,mark=o,mark size=3pt] table [x=param, y=ours, col sep=comma] {figs/eval/various/data/runtime-readpct.csv};
      \addplot[color=brown,mark=square,mark size=3pt] table [x=param, y=baseline-pruning, col sep=comma] {figs/eval/various/data/runtime-readpct.csv};
			\addplot[color=violet,mark=triangle,mark size=3pt] table [x=param, y=baseline, col sep=comma] {figs/eval/various/data/runtime-readpct.csv};
    \end{axis}
		\end{tikzpicture}
		\hspace{1ex}
	\end{scaletikzpicturetowidth}

	\begin{scaletikzpicturetowidth}{0.27\textwidth}
		\begin{tikzpicture}[scale=\tikzscale]
			\begin{axis}[
				font=\LARGE,
				title={(f)},
				xlabel={duplicate key proportion (\%)},
				xtick={0,20,40,60,80,100},
				ymax=10,
				legend pos=north west,
			]
      \addplot[color=blue,mark=o,mark size=3pt] table [x=param, y=ours, col sep=comma] {figs/eval/various/data/runtime-dupkeypct.csv};
      \addplot[color=brown,mark=square,mark size=3pt] table [x=param, y=baseline-pruning, col sep=comma] {figs/eval/various/data/runtime-dupkeypct.csv};
			\addplot[color=violet,mark=triangle,mark size=3pt] table [x=param, y=baseline, col sep=comma] {figs/eval/various/data/runtime-dupkeypct.csv};
    \end{axis}
		\end{tikzpicture}
	\end{scaletikzpicturetowidth}
	}
		\captionsetup{skip=2pt}
  \caption{Verification time comparison  across a range of general workloads.
	Timeout (60s) data points  are not plotted.}
	\label{fig:various-runtime}
\vspace{-2ex}
\end{figure}

%% file: figs/eval/uv/runtime/runtime.tex

\pgfplotsset{height=140pt, width=240pt}
\pgfplotsset{tick style={draw=none}}
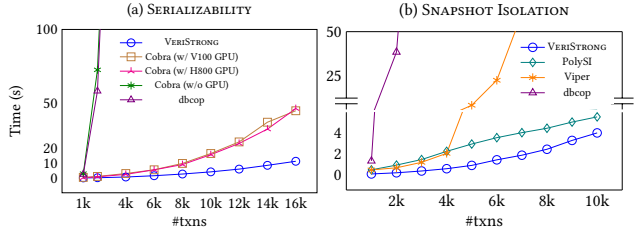
\begin{figure}[t]
	\centering
	\resizebox{\linewidth}{!}{
	\begin{minipage}{0.31\textwidth}
		\begin{scaletikzpicturetowidth}{\textwidth}
			\begin{tikzpicture}[scale=\tikzscale]
				\begin{axis}[
				font=\LARGE,
					title={(a) \ser},
					xlabel={\#txns},
					xtick={1000,4000,6000,8000,10000,12000,14000,16000},
					xticklabels={1k,4k,6k,8k,10k,12k,14k,16k},
					xticklabel style={/pgf/number format/fixed},
					scaled x ticks=false,
					ylabel={Time (s)},
					ymax=100,
					ytick={0,10,20,50,100},
					cycle multiindex* list={
									color       \nextlist
									mark list*  \nextlist
							},
				legend pos=north west,
                legend style={xshift=35pt, yshift=1pt,draw=none},
					height=170pt,
					]
					\addplot[color=blue,mark=o,mark size=3pt] table [x=param, y=ours, col sep=comma] {figs/eval/uv/runtime/ser.csv};
					\addplot[color=brown,mark=square,mark size=3pt] table [x=param, y=cobra(w/v100), col sep=comma] {figs/eval/uv/runtime/ser.csv};
					\addplot[color=magenta,mark=Mercedes star,mark size=3pt] table [x=param, y=cobra(w/h800), col sep=comma] {figs/eval/uv/runtime/ser.csv};
					\addplot[color=green!50!black,mark=asterisk,mark size=3pt] table [x=param, y=cobra(w/ogpu), col sep=comma] {figs/eval/uv/runtime/ser.csv};
					\addplot[color=violet,mark=triangle,mark size=3pt] table [x=param, y=dbcop, col sep=comma] {figs/eval/uv/runtime/ser.csv};
					\legend{\small \ourtool, \small \cobra(w/ V100 GPU), \small \cobra(w/ H800 GPU), \small \cobra(w/o GPU), \small \dbcop}
				\end{axis}
			\end{tikzpicture}
		\end{scaletikzpicturetowidth}
	\end{minipage}

	\hfil
	\begin{minipage}{0.31\textwidth}
		\begin{scaletikzpicturetowidth}{\textwidth}
			\begin{tikzpicture}[scale=\tikzscale]
				\begin{groupplot}[
					group style={
						group size=1 by 2,
						vertical sep=2pt,
					},
					xmin=0, xmax=11000,
					xtick={2000,4000,6000,8000,10000},
					xticklabels={2k,4k,6k,8k,10k},
					scaled x ticks=false,
					cycle multiindex* list={
						color       \nextlist
						mark list*  \nextlist
					},
				legend pos=north east,
				legend style={yshift=-2pt,draw=none},
				]
					\nextgroupplot[
				font=\LARGE,
						title={(b) \si},
            ymin=6, ymax=50, 
						ytick={25,50},
						axis y discontinuity=parallel,
						xticklabel=\empty, 
						height=100pt,
						axis x line=top, 
						axis line style={-},     
						xtick align=inside,
          ]
					\addplot[color=blue,mark=o,mark size=3pt] table [x=param, y=ours, col sep=comma] {figs/eval/uv/runtime/si-raw.csv};
					\addplot[color=teal,mark=diamond,mark size=3pt] table [x=param, y=polysi, col sep=comma] {figs/eval/uv/runtime/si-raw.csv};
					\addplot[color=orange,mark=asterisk,mark size=3pt] table [x=param, y=viper, col sep=comma] {figs/eval/uv/runtime/si-raw.csv};
					\addplot[color=violet,mark=triangle,mark size=3pt] table [x=param, y=dbcop, col sep=comma] {figs/eval/uv/runtime/si-raw.csv};
					\legend{\small \ourtool, \small  \polysi, \small  \viper, \small  \dbcop};

          \nextgroupplot[
				font=\LARGE,
            ymin=-1.5, ymax=6, 
						xlabel={\#txns},
						ytick={0,2,4},
						axis x line*=none, 
						height=100pt,
          ]
					\addplot[color=blue,mark=o,mark size=3pt] table [x=param, y=ours, col sep=comma] {figs/eval/uv/runtime/si-raw.csv};
					\addplot[color=teal,mark=diamond,mark size=3pt] table [x=param, y=polysi, col sep=comma] {figs/eval/uv/runtime/si-raw.csv};
					\addplot[color=orange,mark=asterisk,mark size=3pt] table [x=param, y=viper, col sep=comma] {figs/eval/uv/runtime/si-raw.csv};
					\addplot[color=violet,mark=triangle,mark size=3pt] table [x=param, y=dbcop, col sep=comma] {figs/eval/uv/runtime/si-raw.csv};
				\end{groupplot}
			\end{tikzpicture}

		\end{scaletikzpicturetowidth}
	\end{minipage}
	}
		\captionsetup{skip=0pt}
	\caption{Comparison with existing verifiers under \uniqueval{} histories using
the BlindW-RW datasets  (with 10k keys in (a) and 2k keys in (b)).}
	
	\label{fig:uv-runtime}
\vspace{-2ex}
\end{figure}

%% file: figs/eval/scalability/scala.tex

\pgfplotsset{height=180pt, width=240pt}
\pgfplotsset{tick style={draw=none}}
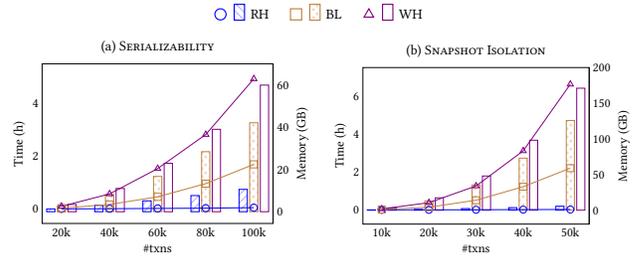
\begin{figure}
	\centering
	\resizebox{.4\linewidth}{!}{
		\input{figs/eval/scalability/legend}
	}

	\resizebox{\linewidth}{!}{
	\begin{scaletikzpicturetowidth}{0.4\columnwidth}
		\begin{tikzpicture}[scale=\tikzscale]
			\begin{axis}[
				font=\Large,
				ymax=70,
				ybar,
				bar width=7pt,
				axis y line*=right,
				axis x line=none,
				ylabel={Memory (GB)},
				ylabel style={rotate=-90,anchor=south},
				ylabel near ticks,
				ymajorgrids=false,
				legend pos=north west,
				legend style={yshift=-12pt},
        xtick={20, 40, 60, 80, 100},
			]
			\addplot+[pattern color=blue!30,pattern=north east lines,draw=blue] table [x=param, y=grh, col sep=comma] {figs/eval/scalability/txns-ser/mem.csv};
			\addplot+[pattern color=brown!30,pattern=crosshatch dots,draw=brown] table [x=param, y=grw, col sep=comma] {figs/eval/scalability/txns-ser/mem.csv};
			\addplot+[pattern color=violet!30,pattern=none,draw=violet] table [x=param, y=gwh, col sep=comma] {figs/eval/scalability/txns-ser/mem.csv};
			\end{axis}

			\begin{axis}[
				font=\Large,
				title={\LARGE (a) \ser},
				xlabel={\#txns},
				ylabel={Time (h)},
				ylabel near ticks,
				ymax=5.5,
				legend pos=north west,
				legend style={yshift=20pt,legend columns=3},
        xtick={20, 40, 60, 80, 100},
				xticklabels={},
				xticklabels={20k, 40k, 60k, 80k, 100k},
			]
				\addplot[color=blue,mark=o,mark size=3pt] table [x=param, y=grh, col sep=comma] {figs/eval/scalability/txns-ser/runtime.csv};
				\addplot[color=brown,mark=square,mark size=3pt] table [x=param, y=grw, col sep=comma] {figs/eval/scalability/txns-ser/runtime.csv};
				\addplot[color=violet,mark=triangle,mark size=3pt] table [x=param, y=gwh, col sep=comma] {figs/eval/scalability/txns-ser/runtime.csv};
				
			\end{axis}
		\end{tikzpicture}
	\end{scaletikzpicturetowidth}
		\hspace{1ex}
	\begin{scaletikzpicturetowidth}{0.4\columnwidth}
		\begin{tikzpicture}[scale=\tikzscale]
			\begin{axis}[
				font=\Large,
				ymax=200,
				ybar,
				bar width=7pt,
				axis y line*=right,
				axis x line=none,
				ylabel={Memory (GB)},
				ylabel style={rotate=-90,anchor=south},
				ylabel near ticks,
				ymajorgrids=false,
				legend pos=north west,
				legend style={yshift=-12pt},
        xtick={10, 20, 30, 40, 50},
			]
			\addplot+[pattern color=blue!30,pattern=north east lines,draw=blue] table [x=param, y=grh, col sep=comma] {figs/eval/scalability/txns-si/mem.csv};
			\addplot+[pattern color=brown!30,pattern=crosshatch dots,draw=brown] table [x=param, y=grw, col sep=comma] {figs/eval/scalability/txns-si/mem.csv};
			\addplot+[pattern color=violet!30,pattern=none,draw=violet] table [x=param, y=gwh, col sep=comma] {figs/eval/scalability/txns-si/mem.csv};
			\end{axis}

			\begin{axis}[
				font=\Large,
				title={\LARGE (b) \si},
				xlabel={\#txns},
				ylabel={Time (h)},
				ylabel near ticks,
				ymax=7.5,
				legend pos=north west,
				legend style={yshift=20pt,legend columns=3},
        xtick={10, 20, 30, 40, 50},
				xticklabels={},
				xticklabels={10k, 20k, 30k, 40k, 50k},
			]
				\addplot[color=blue,mark=o,mark size=3pt] table [x=param, y=grh, col sep=comma] {figs/eval/scalability/txns-si/runtime.csv};
				\addplot[color=brown,mark=square,mark size=3pt] table [x=param, y=grw, col sep=comma] {figs/eval/scalability/txns-si/runtime.csv};
				\addplot[color=violet,mark=triangle,mark size=3pt] table [x=param, y=gwh, col sep=comma] {figs/eval/scalability/txns-si/runtime.csv};
				
			\end{axis}
		\end{tikzpicture}

	\end{scaletikzpicturetowidth}

	}
		\captionsetup{skip=2pt}
  \caption{\ourtool's performance under  large workloads.
	Time and memory usage are plotted in lines and bars.
	}
	\label{fig:scalability}
\vspace{-2ex}
\end{figure}

%% file: figs/eval/scalability/legend.tex

 \pgfplotsset{%
rh legend/.style={legend image code/.code={%
\node[##1,anchor=west] at (0cm,0cm){\pgfuseplotmark{o}};
\path[#1](0.25cm,-0.05cm)rectangle(.38cm,.15cm);
}},
bl legend/.style={legend image code/.code={%
\node[##1,anchor=west] at (0cm,0cm){\pgfuseplotmark{square}};
\path[#1](0.25cm,-0.05cm)rectangle(.38cm,.15cm);
}},
wh legend/.style={legend image code/.code={%
\node[##1,anchor=west] at (0cm,0cm){\pgfuseplotmark{triangle}};
\path[#1](0.25cm,-0.05cm)rectangle(.38cm,.15cm);
}},
 }

\begin{tikzpicture}
\centering
  \begin{axis}[
      hide axis,
      width=2cm, 
      height=2cm, 
      legend style={at={(0.5,-0.1)},anchor=south, legend columns=3, draw=none},
      /tikz/every even column/.append style={column sep=0.2cm}
  ]
  \addlegendimage{color=blue,pattern color=blue!30,pattern=north west lines, rh legend}
  \addlegendentry{\tiny RH}
  \addlegendimage{color=brown,pattern color=brown!30,pattern=crosshatch dots,
    bl legend}
  \addlegendentry{\tiny BL}
  \addlegendimage{color=violet, fill=none, wh legend}
  \addlegendentry{\tiny WH}
  \addplot[draw=none] coordinates {(0,0)}; 
  \end{axis}
\end{tikzpicture}

%% file: sections/eval-dissecting.tex

\subsection{Dissecting \ourtool}
\label{subsec:decompose}

In this section, we take a closer look at \ourtool, examining how its individual components and optimization strategies contribute to the overall checking efficiency.

\input{figs/eval/decomposition/decomposition.tex}

\input{figs/eval/new-ablation/ablation.tex}

\subsubsection{Decomposition Analysis}
We decompose \ourtool's checking time into four stages: constructing, pruning, encoding, and solving.
Figure~\ref{fig:decomposition} presents the breakdown across seven benchmarks (see Section~\ref{ss:setup}).
Overall, the constructing and pruning stages are relatively inexpensive.
The G-TPCC workload is a special case where all constraints are pruned prior to encoding and solving, resulting in negligible cost for the latter stages.
In most benchmarks---except for G-RUBiS and HD---the solving time remains relatively low, whereas the encoding stage dominates.
This behavior is expected: our small-width cycles optimization shifts some of the burden from solving to pruning and encoding, while keeping these two stages efficient.
As a result, the overall checking time is reduced.

\begin{table*}[h]
	\begin{minipage}[t]{0.31\linewidth}
		  \centering
		\pgfplotsset{height=100pt, width=100pt}
		\pgfplotsset{tick style={draw=none}}
	\begin{tikzpicture}
		\begin{axis}[
			tick label style={font=\scriptsize},  
			title={},
			xlabel={ \small $w$},
			ylabel={\small Time (s)},
			ymax=1000,
			ymode=log,
			ylabel near ticks,
			xlabel near ticks,
			xtick={1,2,3,4},
			cycle multiindex* list={
				color       \nextlist
				mark list*  \nextlist
			},
			legend style={
				draw=none,
				font={\footnotesize},
				legend image post style={scale=0.8},
				at={(1.2,1.1)}, 
				anchor=north west, 
				legend columns=1 
			},
			]
			\addplot[color=blue,mark=o,mark size=2pt] table [x=param, y=grubis, col sep=comma] {figs/eval/encode-cycle-width/runtime-large-to.csv};
			\addplot[color=brown,mark=square,mark size=2pt] table [x=param, y=gtwitter, col sep=comma] {figs/eval/encode-cycle-width/runtime-large-to.csv};
			\addplot[color=violet,mark=triangle,mark size=2pt] table [x=param, y=grh, col sep=comma] {figs/eval/encode-cycle-width/runtime-large-to.csv};
			\addplot[color=green!50!black,mark=diamond,mark size=2pt] table [x=param, y=grw, col sep=comma] {figs/eval/encode-cycle-width/runtime-large-to.csv};
			\addplot[color=teal,mark=pentagon,mark size=2pt] table [x=param, y=gwh, col sep=comma] {figs/eval/encode-cycle-width/runtime-large-to.csv};
			\addplot[color=orange,mark=asterisk,mark size=2pt] table [x=param, y=gdv, col sep=comma] {figs/eval/encode-cycle-width/runtime-large-to.csv};
			
			
			
			\legend{G-RBUiS,   G-Twitter,  RH,  RW,  WH,  HD}
		\end{axis}
	\end{tikzpicture}
\captionsetup{skip=6pt}
\captionof{figure}{Runtime of Variant (i) under varying encoded cycle widths ($w$). Timeout (600s) data points are not plotted. }
\label{fig:encode-cycle-width}
	\end{minipage}\hfill
		\begin{minipage}[b]{0.31\linewidth}
		\centering
	\centering
\captionsetup{skip=2pt}
\caption{The number of conflicts encountered during solving with or without Optimization \textbf{H}.  } \label{table:conflicts-polarity}
\footnotesize
\begin{tabular}{ccc}
	\toprule
	Benchmark  & \ourtool & w/o \textbf{H} \\
	\midrule
	G-TPCC		 & 0    & 0    \\
	G-Twitter  & 79   & 99   \\
	G-RUBiS    & 89   & 246  \\
	RH         & 0    & 0    \\
	BL         & 50   & 497  \\
	WH         & 86   & 1427 \\
	HD         & 936  & 2047 \\
	\bottomrule
\end{tabular}
	\end{minipage}
	\hfill
	\begin{minipage}[b]{0.31\linewidth}
  \centering
\captionsetup{skip=6pt}
\caption{Average checking time on anomalous histories (excluding those that timed out beyond 900s). }
\footnotesize
\begin{tabular}{ccc}
	\toprule
	{Verifier} & 
	{SER} & 
	{SI} \\
	\midrule
	Cobra      &   1293ms      &     --        \\ 
	PolySI     &     --           &   359ms     \\
	Viper      &       --       &     6559ms    \\
	dbcop      &   551ms       &   108ms     \\
	\ourtool   &   143ms       &   18ms      \\
	\midrule
	{\#histories} & 2073 & 434 \\
	\bottomrule
\end{tabular}
\label{table:reproduce-violations}
	\end{minipage}
\vspace{-2ex}
\end{table*}

\subsubsection{Ablation Analysis}
To assess the contributions of the major optimizations in  \ourtool, we consider three variants:
(i) \ourtool without the heuristic for polarity picking (\textbf{H});
(ii)  \ourtool without  \textbf{H} and without the encoding of small-width cycles (\textbf{C}); and
(iii) \ourtool without \textbf{H}, \textbf{C}, and pruning (\textbf{P}).

Figure~\ref{fig:ablation} presents our ablation analysis results.
Overall, all optimizations proves effective, though their impact varies across different workloads.
Optimization \textbf{H}
is particularly beneficial for G-RUBiS, BL, WH, and HD, but shows little effect on G-Twitter and RH.
This is because the latter two  are read-heavy workloads, where cycles are rarely encountered during solving.
Consequently, the potential for \textbf{H}  to resolve conflicts through polarity picking is limited.

We also observe that the workloads where Optimization \textbf{C}  is most effective largely overlap with those where \textbf{H}  is effective.
Interestingly, for RH, Optimization \textbf{C} slightly increases the overall checking time.
This is expected, as the overhead stems from the enumeration of small-width cycles, which might be unnecessary in cycle-sparse workloads like RH.


Finally, we find that Optimization \textbf{P} provides significant gains, 
as a large number of 1-width cycles are eliminated,
 including $\WR$ constraints and the associated $\RW$ ones unique to \duplicateval{} histories. 
For instance,
as noted earlier, all constraints in G-TPCC are pruned, allowing the solving phase to be bypassed entirely.

\subsubsection{Impact of  Cycle Width}
\label{sss:choosing-encode-width}
Recall that in Optimization  \textbf{C}, we encode 2-width cycles
(see Section~\ref{sss:encode-2-width-cycles}).
A natural question arises: what is the performance impact of encoding cycles of larger widths?
To answer this, we conduct a comparative experiment using variants of Variant (i), varying the width of encoded cycles.
We plot the results in Figure~\ref{fig:encode-cycle-width},
where $w=1$ corresponds to a baseline with pruning only
(i.e., no cycle encoding),
while $w \geqslant 2$ indicates that cycles of width $w$ are explicitly encoded.
Across all six benchmarks (excluding G-TPCC, where solving is bypassed),
encoding 2-width cycles ($w=2$) consistently delivers the best performance,
striking a balance between pruning effectiveness and solving overhead.
The only exception is RH, where solving time is already low,
suggesting that in cycle-sparse workloads,
the modest overhead of encoding small-width cycles has limited impact
on overall performance.

\subsubsection{Impact of Polarity Picking}
\label{subsubsec:impact-polarity}

To better understand how Optimization \textbf{H} improves \ourtool's performance,
we examine its effect on reducing the number of encountered conflicts (i.e., the number of backtracks) during solving.
We conduct a comparative experiment between \ourtool and Variant (i) across seven benchmarks.
As shown in Table~\ref{table:conflicts-polarity},
 \textbf{H} significantly reduces the number of conflicts.
This effect becomes more pronounced as the write proportion
and the duplication rate of write values increase,
which is consistent with the performance differences observed in Figure~\ref{fig:ablation}.
The only two exceptions are G-TPCC and RH, both of which exhibit no conflicts.
In G-TPCC, all constraints are pruned prior to solving.
In RH, although not all constraints are pruned,
its read-heavy nature results in conflict-free solving even without \textbf{H}.

%% file: figs/eval/decomposition/decomposition.tex

\pgfplotsset{tick style={draw=none}}
\begin{figure}[t]
  \centering

    \begin{scaletikzpicturetowidth}{\columnwidth}
      \begin{tikzpicture}[scale=\tikzscale]
        \begin{groupplot}[
          group style={
            group size=1 by 2, 
            vertical sep=0cm,
            xticklabels at=edge bottom,
          },
          width=12cm,
          height=3.2cm,
          x tick style={draw=none},
          xmin=-0.5,
          xmax=6.5,
          xtick=data,
          xticklabels={G-TPCC,G-Twitter,G-RUBiS,RH,BL,WH,HD},
          ybar stacked,
          ylabel near ticks,
          area legend,
          legend style={at={(0.03, 0.8)}, anchor=west,legend columns=4,draw=none,column sep=4pt}
        ]
          \nextgroupplot[
            font=\small,
            ymin=5, ymax=100, 
            ytick={50,100},
            axis y discontinuity=parallel,
            axis x line=bottom,
            x axis line style ={dotted, -},
            axis y line=left,
            ylabel={Time (s)},
            ylabel style={xshift=-1.2cm},
          ]
          \addplot[color=ps1green, fill=white] table [x expr=\coordindex, y=construct, col sep=comma] {figs/eval/decomposition/data2.csv};
          \addplot[color=ps3green, pattern color=ps3green, pattern=north east lines] table [x expr=\coordindex, y=prune, col sep=comma] {figs/eval/decomposition/data2.csv};
          \addplot[color=ps5green, pattern color=ps5green, pattern=crosshatch dots] table [x expr=\coordindex, y=encode, col sep=comma] {figs/eval/decomposition/data2.csv};
          \addplot[color=ps7green, fill=ps7green] table [x expr=\coordindex, y=solve, col sep=comma] {figs/eval/decomposition/data2.csv};
          \legend{\small Constructing, \small Pruning, \small  Encoding, \small Solving}

          \nextgroupplot[
            font=\small,
            ymin=0, ymax=5, 
            axis x line=bottom,
            axis y line=left,
            y axis line style={-},
          ]
          \addplot[color=ps1green, fill=white] table [x expr=\coordindex, y=construct, col sep=comma] {figs/eval/decomposition/data.csv};
          \addplot[color=ps3green, pattern color=ps3green, pattern=north east lines] table [x expr=\coordindex, y=prune, col sep=comma] {figs/eval/decomposition/data.csv};
          \addplot[color=ps5green, pattern color=ps5green, pattern=crosshatch dots] table [x expr=\coordindex, y=encode, col sep=comma] {figs/eval/decomposition/data.csv};
          \addplot[color=ps7green, fill=ps7green] table [x expr=\coordindex, y=solve, col sep=comma] {figs/eval/decomposition/data.csv};
        \end{groupplot}

      \end{tikzpicture}
    \end{scaletikzpicturetowidth}

	\captionsetup{skip=2pt}
  \caption{Breakdown of checking time across stages.}
  \label{fig:decomposition}
	\vspace{-2ex}
\end{figure}

%% file: figs/eval/new-ablation/ablation.tex
\begin{figure}[t]
  \centering
  \hspace{0.05\linewidth}
  \resizebox{.9\linewidth}{!}{
		\input{figs/eval/new-ablation/legend}
	}

  \resizebox{\linewidth}{!}{
  \begin{scaletikzpicturetowidth}{0.48\textwidth}
    \begin{tikzpicture}[scale=\tikzscale]
      \begin{groupplot}[
        group style={
          group size=1 by 3,
          vertical sep=0pt,
          ylabels at=edge left,
          xticklabels at=edge bottom,
          yticklabels at=edge left,
        },
        width=15cm,
        legend style={at={(0.04, 0.5)}, anchor=north west, yshift=0.6cm, font=\normalsize, legend columns=2,draw=none,column sep=6pt},
        x tick style={draw=none},
        ylabel={Time (s) in log scale},
        ybar,
        area legend,
        xmin=-0.5,
        xmax=6.5,
        xtick=data,
        xtick distance=3,
        xticklabels={G-TPCC,G-Twitter,G-RUBiS,RH,BL,WH,HD},
        xlabel style={font=\Large}, 
        				ylabel near ticks,
        ]

        \nextgroupplot[
          font=\large,
          ymin=60, ymax=600,
          ytick={200, 400, 600},
          axis x line=bottom,
          x axis line style={dotted, -},
          axis y line=left,
          ylabel={},
          axis y discontinuity=parallel,
          height=3cm,
          yticklabel style={font=\footnotesize},
          y filter/.expression={y < 60 ? nan : y},
          ylabel={Time (s)},
          ylabel style={xshift=-1cm},
          nodes near coords,
          point meta=explicit symbolic,
          nodes near coords style={font=\scriptsize, anchor=south, yshift=-0.5pt},
          enlarge y limits={value=0.1,upper},
        ]

        \addplot[
          color=blue, 
          pattern color=blue, 
          pattern=none, 
          bar width = 8pt
        ] table [
          x expr=\coordindex, 
          y=ours, 
          col sep=comma, 
          meta=ours-label
        ] {figs/eval/new-ablation/data.csv};
        \addplot[
          color=brown, 
          pattern color=brown, 
          pattern=crosshatch dots, 
          bar width = 8pt
        ] table [
          x expr=\coordindex, 
          y=ours-h, 
          col sep=comma,
          meta=ours-h-label
        ] {figs/eval/new-ablation/data.csv};
        \addplot[
          color=violet, 
          pattern color=violet, 
          pattern=north east lines, 
          bar width = 8pt
        ] table [
          x expr=\coordindex, 
          y=ours-hc, 
          col sep=comma,
          meta=ours-hc-label
        ] {figs/eval/new-ablation/data.csv};
        \addplot[
          color=green!50!black, 
          pattern color=green!50!black, 
          pattern=crosshatch, 
          bar width = 8pt,
        ] table [
          x expr=\coordindex, 
          y=ours-hcp, 
          col sep=comma, 
          meta=ours-hcp-label
        ] {figs/eval/new-ablation/data.csv};



        \nextgroupplot[
          font=\large,
          ymin=5, ymax=60,
          ytick={20, 30, 40, 50, 60},
          axis x line=bottom,
          axis y line=left,
          x axis line style={dotted, -},
          y axis line style={-},
          xlabel={},
          ylabel={},
          height=3cm,
          yticklabel style={font=\footnotesize},
          axis y discontinuity=parallel,
          nodes near coords,
          point meta=explicit symbolic,
          nodes near coords style={font=\scriptsize, anchor=south, yshift=-0.5pt},
          enlarge y limits={value=0.1,upper},
        ]
        
        \addplot[
          color=blue, 
          pattern color=blue, 
          pattern=none, 
          bar width = 8pt
        ] table [
          x expr=\coordindex, 
          y=ours, 
          col sep=comma, 
          meta=ours-label
        ] {figs/eval/new-ablation/data.csv};
        \addplot[
          color=brown, 
          pattern color=brown, 
          pattern=crosshatch dots, 
          bar width = 8pt
        ] table [
          x expr=\coordindex, 
          y=ours-h, 
          col sep=comma,
          meta=ours-h-label
        ] {figs/eval/new-ablation/data.csv};
        \addplot[
          color=violet, 
          pattern color=violet, 
          pattern=north east lines, 
          bar width = 8pt
        ] table [
          x expr=\coordindex, 
          y=ours-hc, 
          col sep=comma,
          meta=ours-hc-label
        ] {figs/eval/new-ablation/data.csv};
        \addplot[
          color=green!50!black, 
          pattern color=green!50!black, 
          pattern=crosshatch, 
          bar width = 8pt,
        ] table [
          x expr=\coordindex, 
          y=ours-hcp, 
          col sep=comma, 
          meta=ours-hcp-label
        ] {figs/eval/new-ablation/data.csv};

        \nextgroupplot[
          font=\large,
          ymin=0, ymax=5,
          ytick={0, 2, 4},
          axis x line=bottom,
          axis y line=left,
          y axis line style={-},
          xlabel={},
          ylabel={},
          height=3cm,
          yticklabel style={font=\footnotesize},
          xticklabel style={font=\large},
          nodes near coords,
          point meta=explicit symbolic,
          nodes near coords style={font=\scriptsize, anchor=south, yshift=-0.5pt},
          enlarge y limits={value=0.1,upper},
        ]

        
        \addplot[
          color=blue, 
          pattern color=blue, 
          pattern=none, 
          bar width = 8pt
        ] table [
          x expr=\coordindex, 
          y=ours, 
          col sep=comma, 
          meta=ours-label
        ] {figs/eval/new-ablation/data.csv};
        \addplot[
          color=brown, 
          pattern color=brown, 
          pattern=crosshatch dots, 
          bar width = 8pt
        ] table [
          x expr=\coordindex, 
          y=ours-h, 
          col sep=comma,
          meta=ours-h-label
        ] {figs/eval/new-ablation/data.csv};
        \addplot[
          color=violet, 
          pattern color=violet, 
          pattern=north east lines, 
          bar width = 8pt
        ] table [
          x expr=\coordindex, 
          y=ours-hc, 
          col sep=comma,
          meta=ours-hc-label
        ] {figs/eval/new-ablation/data.csv};
        \addplot[
          color=green!50!black, 
          pattern color=green!50!black, 
          pattern=crosshatch, 
          bar width = 8pt,
        ] table [
          x expr=\coordindex, 
          y=ours-hcp, 
          col sep=comma, 
          meta=ours-hcp-label
        ] {figs/eval/new-ablation/data.csv};
      \end{groupplot}


    \end{tikzpicture}
  \end{scaletikzpicturetowidth}
  }
  	\captionsetup{skip=2pt}
  \caption{Ablation analysis. 
  Timeout is set to 600s.}
  \label{fig:ablation}
\vspace{-2ex}
\end{figure}

%% file: figs/eval/new-ablation/legend.tex

 \pgfplotsset{%
ours legend/.style={legend image code/.code={%
\path[#1](-0.0cm,-0.05cm)rectangle(.38cm,.12cm);
}},
ours-H legend/.style={legend image code/.code={%
\path[#1](-0.0cm,-0.05cm)rectangle(.38cm,.12cm);
}},
ours-H-C legend/.style={legend image code/.code={%
\path[#1](-0.0cm,-0.05cm)rectangle(.38cm,.12cm);
}},
ours-H-C-P legend2/.style={legend image code/.code={%
\path[#1](-0.0cm,-0.05cm)rectangle(.38cm,.12cm);
}},
 }

\begin{tikzpicture}
\centering
  \begin{axis}[
      hide axis,
      width=2cm, 
      height=2cm, 
      legend style={at={(0.5,-0.1)},anchor=south, shift={(1.0,0)}, legend columns=4, draw=none},
      /tikz/every even column/.append style={column sep=0.2cm}
  ]
  \addlegendimage{color=blue,pattern color=blue,pattern=none, ours legend}
  \addlegendentry{\tiny \ourtool}
  \addlegendimage{color=brown,pattern color=brown,pattern=crosshatch dots, ours-H legend}
  \addlegendentry{\tiny \ourtool w/o \textbf{H}}
  \addlegendimage{color=violet, pattern color=violet,pattern=north east lines, ours-H-C legend}
  \addlegendentry{\tiny \ourtool w/o \textbf{H} and \textbf{C}}
  \addlegendimage{color=green!50!black, pattern color=green!50!black,pattern=crosshatch, ours-H-C-P legend2}
  \addlegendentry{\tiny \ourtool w/o \textbf{H}, \textbf{C} and \textbf{P}}
  \addplot[draw=none] coordinates {(0,0)}; 
  \end{axis}
\end{tikzpicture}

%% file: sections/eval-effectiveness.tex
\subsection{Effectiveness}\label{subsec:bugs}


Beyond checking performance, 
an essential criterion for a verifier is its ability to accurately detect isolation anomalies.
To this end, we validate \ourtool against an extensive set of known anomallies.


\inlsec{Reproducing \uniqueval{}  Anomalies} 
\ourtool  reproduces all known anomalies from a substantial set of 2507 \uniqueval{}  histories: 2073 for \ser and 434 for \si. 
These histories were originally collected  by prior work~\cite{Cobra:OSDI2020, Complexity:OOPSLA2019, PolySI:VLDB2023} from earlier versions of widely-used databases, including CockroachDB and YugabyteDB.
In addition, we report the runtime of each tool: 
as shown in Table~\ref{table:reproduce-violations}, \ourtool achieves significantly lower average checking times than the state-of-the-art verifiers.


\inlsec{(Re)discovering  \duplicateval{} Anomalies}
Motivated by recent bug reports~\cite{MySQL-Bug, MariaDB-Bug}, we apply \ourtool to test both MySQL and MariaDB.
\ourtool replicates the reported anomalies in earlier versions of both databases and, notably, rediscovers the same bug in the latest stable version  of MariaDB (v 11.5.2).
We reported this issue to the developers, who confirmed it as a valid bug.
Note that all these anomalies lie beyond the capabilities of existing verifiers, which are limited to verifying \uniqueval{} histories.



%% file: sections/related-nobi-0424.tex
\section{Related Work}
\label{section:related}

Recent years have seen significant progress in
 verifying 
database isolation guarantees. We focus on the black-box approaches.

A recent line of work~\cite{Cobra:OSDI2020,PolySI:VLDB2023,Viper:EuroSys2023}, including Cobra and PolySI, 
 applies SMT solving
to verify strong isolation.
All these tools  encode  histories as polygraphs
or their variants and rely on the off-the-shelf solver MonoSAT
to detect cycles that represent anomalies.

\ourtool follows this approach but differs in three key ways.
First, rather than relying solely on general-purpose SMT solvers,
it incorporates workload-specific optimizations to enhance performance. 
Second, existing polygraph-based approaches
are limited to \uniqueval{} histories.
\ourtool introduces an expressive hyper-polygraph representation
that also supports \duplicateval{}, thereby broadening applicability.
Third, \ourtool provides 
both soundness and completeness,  ensuring reliable isolation verification. 
In contrast, 
existing verifiers  risk false positives and false negatives,
particularly in the presence of duplicate write values.

Non-solver tools~\cite{AWDIT:PLDI2025,Complexity:OOPSLA2019,Elle:VLDB2020,MiniTransactions:ICDE2025,Plume:OOPSLA2024}
employ graph traversal algorithms, such as DFS or optimized variants,
to verify isolation guarantees.
The dbcop tool~\cite{Complexity:OOPSLA2019}
applies a polynomial-time algorithm to verify \ser
over \uniqueval{} histories (with a bounded number of sessions), 
along with a polynomial-time reduction from verifying \si to verifying \ser.
However, dbcop has been reported to be less efficient
than Cobra~\cite{Cobra:OSDI2020} and PolySI~\cite{PolySI:VLDB2023},
which in turn are outperformed by \ourtool.

Elle~\cite{Elle:VLDB2020}, the isolation checker used in  the Jepsen framework~\cite{Jepsen}, infers version orders (i.e., \WW{} dependencies) from list-append workloads. 
For example, reading the list \texttt{[2,1]} on key $x$ indicates that $A(x,2)$ precedes $A(x,1)$ in the version order. 
However, such inference relies on the \uniqueval{} assumption, which limits its effectiveness when applied to general workloads containing duplicate write values, e.g., when reading a list \texttt{[1,1]}. 
MTC~\cite{MiniTransactions:ICDE2025} exploits a class of simplified database workloads,
called mini-transactions (e.g.,  read-modify-write transactions)
to achieves quadratic-time verification under the same \uniqueval{} assumption.
While MTC strikes a practical balance between performance and bug-finding effectiveness, its applicability is limited, as real-world workloads often diverge from such simplified transaction patterns.

Other tools,
such as Plume~\cite{Plume:OOPSLA2024} and AWDIT~\cite{AWDIT:PLDI2025}, focus on verifying 
weaker isolation guarantees~\cite{hat,noc-noc}.
These properties are 
 computationally less complex than the stronger ones we target~\cite{Complexity:OOPSLA2019}.

%% file: sections/discussion-nobi-0424.tex
\section{Discussion and Conclusion}
\label{section:discuss}

\inlsec{Supporting Other Formalisms}
Our hyper-polygraphs  are not restricted to  Adya's formalism; rather, they are expressive enough to support other dependency-based characterizations of isolation guarantees, such as the axiomatic framework proposed by Biswas and Enea~\cite{Complexity:OOPSLA2019}. 
These frameworks were introduced to capture more recent isolation levels like 
\ra~\cite{RAMP:TODS2016}.
Our approach supports them by enabling precise modeling of both certain and uncertain dependencies, e.g.,  commit order~\cite{Complexity:OOPSLA2019},
 which, like $\WW$, models the version order. 
As a demonstration, we have encoded Biswas and Enea's framework
 using hyper-polygraphs; see Appendix~\ref{section:appendix-enea}.


\inlsec{Supporting Weak Isolation Levels}
While our dedicated SMT solving is highly efficient for verifying strong isolation levels,
it may be less suitable for weaker ones, where SMT   may be unnecessarily heavyweight~\cite{Plume:OOPSLA2024}.
In practice, users can combine \ourtool with existing weak isolation verifiers~\cite{Plume:OOPSLA2024,AWDIT:PLDI2025} to achieve broader coverage.
This integration is straightforward in the black-box setting, as all tools share a unified input format.
We have realized such an integration   with Plume
in the IsoVista system~\cite{IsoVista:VLDB24}.

\inlsec{Limitations and Future Work}
While our work lifts the  strong  \uniqueval{} assumption  made by prior verifiers, it currently targets key-value databases. Extending our approach to relational models
is a promising research direction, e.g., by building on 
techniques from~\cite{10.1145/3728953,TxCheck:OSDI2023}. 
Another  direction is to unify correctness reasoning for both isolation guarantees and query semantics~\cite{258888,10.1145/3749186} across different databases, drawing inspiration from  recent advances like~\cite{10.1145/3720504}. 

\inlsec{Conclusion}
We have presented a novel approach for verifying strong database isolation, centered around hyper-polygraphs---a general formalism for modeling dependency-based isolation semantics. 
We have focused on Adya's theory and established sound and complete characterizations for both \ser and \si.
Our verifier achieves high efficiency by tailoring SMT solving to database  workload characteristics, in contrast to the state-of-the-art that relies solely on general-purpose solvers.




%% file: sections/si-nobi-0424.tex
\section{Verifying \si}
\label{section:appendix-si}

To demonstrate the generality of our proposed techniques, this section summarizes how hyper-polygraphs can be used to characterize \si (SI), a widely used strong isolation guarantee in practice, and 
how our optimizations extend naturally to verifying histories against SI.

Unlike verifying \ser, where detecting any cycle is sufficient to determine a violation, verifying \si is more complex, requiring identifying cycles that do not contain two adjacent $\RW$ edges~\citep[Theorem 4.1]{AnalysingSI:JACM2018}.
Inspired by prior work~\cite{PolySI:VLDB2023}, 
we define the \emph{induced SI graph} based on a 
directed labeled graph that is
 compatible with the hyper-polygraph of a history (as specified in Definition~\ref{def:compatible-graphs-with-a-hyper-polygraph}).

\begin{definition} \label{def:induced-SI-graph}
Let $\G' = (\V', \E')$ be a graph compatible with a hyper-polygraph $\G$.
Let $\E_{\typevar}$ ($\typevar \in \type$) denote the subset of $\E$ that consists of edges labeled with $\typevar$.
The \emph{induced SI graph} of $\G'$ is defined as
\[
\G'_{\textsc{I}} = (\V', (\E'_{\SO} \cup \E'_{\WR} \cup \E'_{\WW}) \comp \E'_{\RW}?).\footnote{
  Here, we slightly abuse the notation of $\comp$ and $?$, where
  $\E_{\typevar}$ is treated as a subset of $\V \times \V$ 
  by selecting edges labeled $\typevar$ in $\E$ and ignoring keys .
}
\]
\end{definition}

We can then characterize \si  using hyper-polygraphs  as follows.
Illustrative examples are given below, and the proof is provided in Section~\ref{ss:proof-hpg-si}.

\begin{theorem}
  \label{thm:si-hpg}
  A history $\H$ satisfies \emph{\si{}} if and only if
  \emph{$\H \models \intaxiom$} and
  there exists a graph compatible with the hyper-polygraph of $\H$ 
  whose induced SI graph is acyclic.
\end{theorem}






%% file: appendix/app-si.tex

The following example illustrates the induced SI graph, 
reusing the examples in Figure~\ref{fig:rw-checking}. 

\input{figs/rw-checking-si-induced}

\begin{example}
  \label{ex:induced-si-graph}
  We reuse the history in Example~\ref{ex:hyper-polygraph} (Figure~\ref{fig:rw-checking}) to illustrate induced SI graphs.
  Figure~\ref{fig:rw-checking-si-induced} shows induced SI graphs of the compatible graphs in Figure~\ref{fig:rw-checking}.
  The edges of induced SI graph of the graph in Figure~\ref{fig:rw-checking-h2-solution} 
  (shown in Figure~\ref{fig:rw-checking-h2-solution-si-induced}) are the same as the compatible graph, 
  since there is no $\RW$ edge.
  In constrast, the induced SI graph of the graph in Figure~\ref{fig:rw-checking-h2-conflict} 
  (shown in Figure~\ref{fig:rw-checking-h2-conflict-si-induced}) consists of two new induced edges (shown as dashed arrows):
  $T_1 \rel{} T_2$ (induced from $T_1 \rel{\WR(x)} T_3$ and $T_3 \rel{\RW(x)} T_2$) and 
  $T_2 \rel{} T_2$ (induced from $T_2 \rel{\SO} T_3$ and $T_3 \rel{\RW(x)} T_2$).
  Note that $T_2 \rel{} T_2$ is actually a self-loop, 
  indicating that the induced SI graph is cyclic.
\end{example}

The high-level workflow for verifying \si is aligned with that for \ser, as illustrated in Figure~\ref{fig:ser-smt-theory-framework}.
The  key difference lies in the theory solver:
for \si, the verification procedure maintains the induced SI graph constructed from the known graph.
Both cycle detection and conflict clause generation are performed on this induced graph. 
Additionally, the small-width cycle optimization is adapted to prune and encode cycles in the induced SI graph, based on the composition rule $(\SO \cup \WR \cup \WW) \comp \RW?$.
The polarity picking heuristic applies straightforwardly in this setting.

%% file: figs/rw-checking-si-induced.tex

\begin{figure}[h]
	  \vspace{-2ex}
  \centering
  \begin{subfigure}[b]{0.23\textwidth}
    \centering
    \includegraphics[width = 0.70\textwidth]{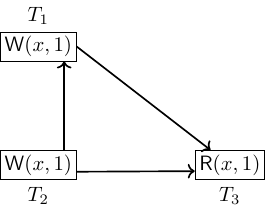}
      \captionsetup{skip=4pt}
    \caption{The induced SI graph of the graph in Figure~\ref{fig:rw-checking-h2-solution}.}
    \label{fig:rw-checking-h2-solution-si-induced}
  \end{subfigure}
  \hfill
  \begin{subfigure}[b]{0.24\textwidth}
    \centering
    \includegraphics[width = 0.8\textwidth]{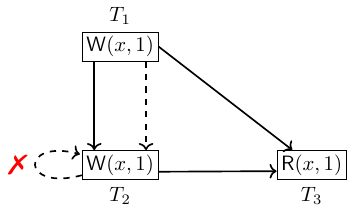}
      \captionsetup{skip=4pt}
    \caption{The induced SI graph of the graph in Figure~\ref{fig:rw-checking-h2-conflict}.}
    \label{fig:rw-checking-h2-conflict-si-induced}
  \end{subfigure}
  \captionsetup{skip=4pt}
  \caption{Induced SI graphs of compatible graphs in Figure~\ref{fig:rw-checking}.}
  \label{fig:rw-checking-si-induced}
  \vspace{-2ex}
\end{figure}

%% file: appendix/app-proof.tex
\section{Proofs}
\label{section:appendix-proofs}

\subsection{Proof of Theorem~\ref{thm:ser-hpg}}
\label{ss:appendix-proof-hpg-ser}

\begin{proof}
  The proof proceeds in two directions.

  (``$\Rightarrow$'')
  Suppose that $\H$ satisfies \ser.
  By Theorem~\ref{thm:ser-depgraph}, $\H$ satisfies \intaxiom{},
  and there exists $\WR$, $\WW$ and $\RW$ relations with which
  $\H$ can be extended to a dependency graph $\G$ such that
  $\SO_\G \cup \WR_\G \cup \WW_\G \cup \RW_\G$ is acyclic.
  Denote the hyper-polygraph of $\H$ as $\G' \triangleq (\V', \E', \C')$.
  We construct an acyclic compatible graph $\G'' \triangleq (\V'', \E'')$ of $\G'$ based on $\G$,
  where $\V'' = \V'$ corresponds to transactions of $\H$ by definition,
  and $\E'' = \SO_\G \cup \WR_\G \cup \WW_\G \cup \RW_\G$ with each edge
  labeled by its relation name.
  $\G''$ is acyclic and satisfies Definition~\ref{def:compatible-graphs-with-a-hyper-polygraph}.

  (``$\Leftarrow$'')
  Suppose that $\H$ satisfies \intaxiom{} and
  there exists an acyclic compatible graph $\G''$ of the hyper-polygraph $\G'$ of $\H$.
  We construct a dependency graph $\G$ based on $\G''$,
  where $\SO_\G, \WR_\G, \WW_\G$ and $\RW_\G$ consist a partition of $\E''$ by their labels.
  For example,
  $\SO_\G = \bigcup\limits_{T \rel{\SO} S} (T, S)$,
  and $\forall x \in \Key.\; \WR_\G(x) = \bigcup\limits_{T \rel{\WR(x)} S} (T, S)$
  (the same for $\WW$ and $\RW$).
  $\G$ is a valid dependency graph by Definition~\ref{def:depgraph}.
  Since $\G''$ is acyclic,
  $\SO_\G \cup \WR_\G \cup \WW_\G \cup \RW_\G$ is also acyclic. 
  $\G$ satisfies \ser by Theorem~\ref{thm:ser-depgraph}.
\end{proof}

\subsection{Proof of Theorem~\ref{thm:si-hpg}}
\label{ss:proof-hpg-si}

The proof of the hyper-polygraph-based characterization for \si closely follows the same strategy as that of \ser. 
Both directions of the construction -- 
deriving a valid dependency graph from a compatible graph and vice versa --
are analogous and can be reused verbatim.

\begin{proof}
  This proof proceeds in two directions.

  (``$\Rightarrow$'')
  Suppose that $\H$ satisfies \si.
  By~\citep[Theorem 4.1]{AnalysingSI:JACM2018}, $\H$ satisfies \intaxiom{}, 
  and there exists $\WR$, $\WW$ and $\RW$ relations with which
  $\H$ can be extended to a dependency graph $\G$ such that
  $(\SO_\G \cup \WR_\G \cup \WW_\G) \comp \RW_\G?$ is acyclic.
  Denote the hyper-polygraph of $\H$ as $\G' \triangleq (\V', \E', \C')$.
  We construct an acyclic compatible graph $\G'' \triangleq (\V'', \E'')$ of $\G'$ based on $\G$,
  where $\V'' = \V'$ corresponds to transactions of $\H$ by definition,
  and $\E'' = \SO_\G \cup \WR_\G \cup \WW_\G \cup \RW_\G$ with each edge
  labeled by its relation name.
  The induced SI graph of $\G''$ is acyclic, 
  and $\G''$ satisfies Definition~\ref{def:compatible-graphs-with-a-hyper-polygraph}.

  (``$\Leftarrow$'')
  Suppose that $\H$ satisfies \intaxiom{} and
  there exists a compatible graph $\G''$ of the hyper-polygraph $\G'$ of $\H$ 
  whose induced SI graph $\G''_{\textsc{I}}$ is acyclic.
  We construct a dependency graph $\G$ based on $\G''$,
  where $\SO_\G, \WR_\G, \WW_\G$ and $\RW_\G$ consist a partition of $\E''$ by their labels.
  For example,
  $\SO_\G = \bigcup\limits_{T \rel{\SO} S} (T, S)$,
  and $\forall x \in \Key.\; \WR_\G(x) = \bigcup\limits_{T \rel{\WR(x)} S} (T, S)$
  (the same for $\WW$ and $\RW$).
  $\G$ is a valid dependency graph by Definition~\ref{def:depgraph}.
  Since $\G''_{\textsc{I}}$ is acyclic,
  $(\SO_\G \cup \WR_\G \cup \WW_\G) \comp \RW_\G?$ is also acyclic. 
  $\G$ satisfies \si by~\citep[Theorem 4.1]{AnalysingSI:JACM2018}.
\
\end{proof}

%% file: appendix/cdclt.tex
\section{An Illustration of CDCL(T)}
\label{section:appendix-cdclt}

\begin{figure}
	\centering
	\includegraphics[width=\columnwidth]{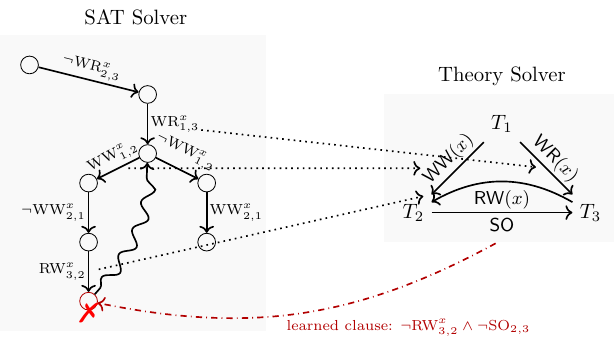}
	\captionsetup{skip=2pt}
	\caption{An illustration of how the SAT solver and the theory solver collaborate in CDCL(T) for the history  $\H$ in Figure~\ref{fig:rw-checking}.}
	\label{fig:CDCL}
\end{figure}


Figure~\ref{fig:CDCL} illustrates the interaction of the SAT solver and the theory solver 
when solving $\H$ in Figure~\ref{fig:rw-checking}. 
The solving procedure is identical to that in Figure~\ref{fig:running-example-solving}; 
see Section~\ref{ss:solving}.

%% file: appendix/app-algo.tex
\section{Other Algorithms}
\label{section:appendix-alg}


Algorithm~\ref{algo:construct} shows the process of constructing the hyper-polygraph $\G$ of a  history $\H$.
It is called at Line~\ref{line:checkser-call-construct} in Algorithm~\ref{alg:solver} 
 (Section~\ref{sec:alg}).


\input{algs/construct.tex}


Algorithm~\ref{algo:fast-prune} shows the process of pruning the hyper-polygraph $\G$.
It is called at Line~\ref{line:checkser-call-prune} in Algorithm~\ref{alg:solver}; 
see Section~\ref{sec:alg}.
\input{algs/fast-prune.tex}

Algorithm~\ref{algo:reachability} shows the process of computing reachability of a given known graph.
It is introduced in the setup part of Section~\ref{section:exp}, 
as an auxiliary procedure in Algorithm~\ref{algo:fast-prune} (\textsc{Prune}) and 
Algorithm~\ref{algo:init-pair-conflict} (\textsc{EncodeCyclesOfWidth2}) 
in acyclicity testing.

\input{algs/reachability.tex}

Algorithm~\ref{algo:encode} shows the process of encoding.
It is called at Line~\ref{line:checkser-call-encode} in Algorithm~\ref{alg:solver}; 
see Section~\ref{sec:alg}.

\input{algs/sat-encode.tex}

To generate conflict clauses, 
we introduce a boolean expression \emph{reason} for each edge in $\ug$.
Let $\mathbb{F}$ denote the set of all boolean formulas.
Formally, we define $\ug \triangleq (\V, \eug)$, where $\eug \subseteq \V \times \V \times \mathbb{F}$.
We write $T \rel{\reasonvar} S$ to denote an edge $(T, S, \reasonvar)$ in $\ug$, 
where $\reasonvar \in \mathbb{F}$ is the reason for the edge from $T$ to $S$.
Edges in the initial known graph are assigned the reason $\true$ 
(Line~\ref{line:solver-init-add-known-graph} in Algorithm~\ref{alg:solver}).

Algorithm~\ref{algo:rw-derive} (\textsc{DeriveRWEdges}) is called at Line~\ref{line:theorysolve-call-derive-rw-edges} in Algorithm~\ref{alg:solver}.
It shows the procedure of generating $\cedges$, 
which includes the $\WW$ (or $\WR$) edge corresponding to the assigned boolean variable 
and the derived $\RW$ edges.
The reason of a $\WW$ or $\WR$ edge is the corresponding boolean variable.
The reason of an $\RW$ edge is the conjunction of the reasons of the $\WW$ edge and the $\WR$ edge 
from where it is derived  
(see Line~\ref{line:derive-rw-ww-case} and Line~\ref{line:derive-rw-wr-case}).

Algorithm~\ref{algo:gen-conflict-clause} (\textsc{GenConflictClause}) is called at 
Line~\ref{line:theorysolve-generate-conflict-clause} in Algorithm~\ref{alg:solver}.
Given a detected cycle, 
the conflict clause is the negation of 
the conjunction of the reasons of the edges involved in the cycle 
(see Line~\ref{line:conflict-clause}).

\input{algs/rw-derive.tex}

\input{algs/gen-conflict-clause.tex}

%% file: algs/construct.tex

\begin{algorithm}
  \small
  \caption{Construct the hyper-polygraph for a history $\H$}
  \label{algo:construct}
  \begin{algorithmic}[1]
    \Statex $\G = (\V, \E, \C)$: the known graph; initially $(\T, \emptyset, (\emptyset, \emptyset))$

    \hStatex
    \Function{Construct}{$\hist$}
      \ForAll{$T, S \in \T$ such that $T \rel{\SO} S$}
        \State $\E \gets \E \;\cup\; \set{T \rel{\SO} S}$
      \EndFor
      \ForAll{$S \in \T$ such that $S \vdash \readevent(x, v)
        \land |\set{T \in \T \mid T \vdash \writeevent(x, v)}| = 1$}
        \State $\E \gets \E \;\cup\; \set{T \rel{\WR(x)} S}$
      \EndFor

      \State $\wwcons = \Bset{\btuple{T \rel{\WW(x)} S, S \rel{\WW(x)} T}
      \mid T \in \WriteTx_{x} \land S \in \WriteTx_{x} \land T \neq S}$

      \State $\wrcons = \Bset{\btuple{\bigcup\limits_{T_{i} \vdash \writeevent(x, v)} \set{T_{i} \rel{\WR(x)} S}}
      \mid S \vdash \readevent(x, v)}$
    \EndFunction
  \end{algorithmic}
\end{algorithm}

%% file: algs/fast-prune.tex
\begin{algorithm}
  \small
  \caption{Prune}
  \label{algo:fast-prune}
  \begin{algorithmic}[1]
    \Function{Prune}{$\G$}
      \Repeat
        \ForAll{$\ccons \gets \set{\eithervar \triangleq T \rel{\WW(x)} S,
          \orvar \triangleq S \rel{\WW(x)} T} \in \wwcons$}
          \State $\eitheredges \gets \set{\eithervar} \cup
            \set{S' \rel{\RW(x)} S \mid T \rel{\WR(x)} S' \in \E}$
          \State $\oredges \gets \set{\orvar} \cup
            \set{T' \rel{\RW(x)} T \mid S \rel{\WR(x)} T' \in \E}$
          \If{$\E \cup \eitheredges$ is cyclic} \label{line:prune-check-cyclicity-1}
            \State $\ccons \gets \consvar \setminus \set{\eithervar}$
            \State{$\wwcons \gets \wwcons \setminus \set{\ccons}$}
          \EndIf
          \If{$\E \cup \oredges$ is cyclic} \label{line:prune-check-cyclicity-2}
            \State $\ccons \gets \consvar \setminus \set{\orvar}$
            \State{$\wwcons \gets \wwcons \setminus \set{\ccons}$}
          \EndIf
          \If{$\ccons = \emptyset$}
            \State{\Return $\false$}
          \EndIf
          \If{$\ccons = \set{\eithervar}$}
            \State $\E \gets \E \cup \eitheredges$
          \EndIf
          \If{$\ccons = \set{\orvar}$}
            \State $\E \gets \E \cup \oredges$
          \EndIf
        \EndFor
        \ForAll{$\ccons \in \wrcons$}
          \ForAll{$\wrvar \gets T \rel{\WR(x)} S \in \ccons$}
            \State $\wredges \gets \set{\wrvar} \cup
              \set{S \rel{\RW(x)} T' \mid T \rel{\WW(x)} T'}$
            \If{$\E \cup \wredges$ is cyclic} \label{line:prune-check-cyclicity-3}
              \State $\ccons \gets \ccons \setminus \set{\wrvar}$
              \State $\wrcons \gets \wrcons \setminus \set{\ccons}$
            \EndIf
          \EndFor
          \If{$\ccons = \emptyset$}
            \State \Return \false
          \EndIf
          \If{$|\ccons| = 1 \land |\ccons| = \set{\wrvar \triangleq T \rel{\WR(x)} S}$}
            \State $\wredges \gets \set{\wrvar} \cup
              \set{S \rel{\RW(x)} T' \mid T \rel{\WW(x)} T'}$
            \State $\E \gets \E \cup \wredges$
          \EndIf
        \EndFor
      \Until{$\G$ remains unchanged}
      \State{\Return $\true$}
    \EndFunction
  \end{algorithmic}
\end{algorithm}

%% file: algs/reachability.tex

\begin{algorithm}
  \small
  \caption{Compute reachability of the known graph}
  \label{algo:reachability}
  \begin{algorithmic}[1]
    \Function{Reachability}{$\V, \E$}
      \State{$\creachability \gets 0^{\cardinality{\V} \times \cardinality{\V}}$}
      \State{$\mathit{reversed\_topo\_order} \gets $ \Call{TopologicalSort}{$\V, \E$}}
      \ForAll{$x \in \mathit{reversed\_topo\_order}$}
        \State{$\creachability_{xx} \gets 1$}
        \ForAll{$(x, y, \_, \_) \in \E$}
          \State{$\creachability_x \gets \creachability_x \cup \creachability_y$}
        \EndFor
      \EndFor
    \EndFunction
  \end{algorithmic}
\end{algorithm}

%% file: algs/sat-encode.tex
\begin{algorithm}
  \small
  \caption{Encode}
  \label{algo:encode}
  \begin{algorithmic}[1]
    \Function{Encode}{$\V, \E, \C$}
      \ForAll{$\set{T \rel{\WW(x)} S, S \rel{\WW(x)} T} \in \wwcons$} \Comment{\small encode $\WW$ constraints}
        \State{$\cvars \gets \cvars \cup \set{\cwwvar_{T, S}^x, \cwwvar_{S, T}^x}$}
        \State{$\cclauses \gets \cclauses \cup \set{\cwwvar_{T, S}^x \lor \cwwvar_{S, T}^x, 
                  \lnot \cwwvar_{T, S}^x \lor \lnot \cwwvar_{S, T}^x} $}
      \EndFor

      \ForAll{$\ccons \in \wrcons$} \Comment{\small encode $\WR$ constraints}
        \State{$\cvars \gets \cvars \cup \set{\cwrvar{x}_{T, S} \mid T \rel{\WR(x)} S \in \ccons}$}
        \State{$\cclauses \gets \cclauses \cup \set{\bigvee\limits_{T \rel{\WR(x)} S \in \ccons}\cwrvar{x}_{T, S}}$}
      \EndFor

      \Statex{}
      \State{\Call{EncodeCyclesOfWidth2}{$\V, \E$}} \Comment{\small Algorithm~\ref{algo:init-pair-conflict}}
    \EndFunction
  \end{algorithmic}
\end{algorithm}

%% file: algs/rw-derive.tex
\begin{algorithm}
  \small
  \caption{Derive $\RW$ edges}
  \label{algo:rw-derive}
  \begin{algorithmic}[1]
    \Function{DeriveRWEdges}{$\pa$}
      \State{$\cedges \gets \emptyset$}
      \ForAll{$\cwwvar_{T, T'}^{x} \in \pa$} \Comment{$\WW$ variables that are assigned \true }
        \State{$\cedges \gets \cedges \cup \set{T \rel{\cwwvar_{T, T'}^x} T'} \cup 
                \set{S \rel{\cwwvar_{T, T'}^x} T' \mid T \rel{\WR(x)} S \in \E} \cup
                \set{S \rel{\cwwvar_{T, T'}^x \land \cwrvar{x}_{T, S}} T' \mid T \rel{\cwrvar{x}_{T, S}} S \in \eug}$}
          \label{line:derive-rw-ww-case}
      \EndFor
      \ForAll{$\cwrvar{x}_{T, S} \in \pa$} \Comment{$\WR$ variables that are assigned \true }
        \State{$\cedges \gets \cedges \cup \set{T \rel{\cwrvar{x}_{T, S}} S} \cup 
                \set{S \rel{\cwrvar{x}_{T, S}} T' \mid T \rel{\WW(x)} T' \in \E} \cup
                \set{S \rel{\cwwvar_{T, T'}^x \land \cwrvar{x}_{T, S}} T' \mid T \rel{\cwwvar_{T, T'}^x} T' \in \eug}$}
          \label{line:derive-rw-wr-case}
      \EndFor
      \State{\Return $\cedges$}
    \EndFunction


  \end{algorithmic}
\end{algorithm}

%% file: algs/gen-conflict-clause.tex
\begin{algorithm}[h]
  \small
  \caption{Generate conflict clause}
  \label{algo:gen-conflict-clause}
  \begin{algorithmic}[1]
    \Function{GenConflictClause}{$\eug, \cedges$}
      \State{$\ccycle \triangleq \set{T_1 \rel{\reasonvar_1} T_2, T_2 \rel{\reasonvar_2} T_3,  \cdots,  T_n \rel{\reasonvar_n} T_1} \gets $ \Call{Cycle}{$\eug \cup \cedges$}} 
      \State{$\cconflictcl \gets \bigvee\limits_{T_i \rel{\reasonvar_i} T_{i + 1} \in \ccycle} (\lnot \reasonvar_i)$}
        \label{line:conflict-clause}
      \State{\Return $\cconflictcl$}
    \EndFunction
  \end{algorithmic}
\end{algorithm}

%% file: appendix/app-correctness.tex

\section{Correctness of \ourtool}
\label{section:appendix-correctness}

In this section, we establish the correctness  of \ourtool with respect to \ser through its
individual optimizations.

\subsection{Small-Width Cycle Preprocessing}
\label{ss:appendix-correctness-cycle}

\inlsec{Eliminating 1-width Cycles}
We formalize the procedure of eliminating 1-width cycles 
(i.e., pruning) using six transition rules on states of form $\tuple{\E, \wwcons, \wrcons}$, 
as shown in Figure~\ref{fig:rules-of-pruning}.

Pruning starts from the state 
$\tuple{\E, \wwcons, \wrcons}$, 
where $\E, \wwcons$ and $\wrcons$
are components of the constructed hyper-polygraph 
$\G = (\V, \E, (\wwcons, \wrcons))$
(see Algorithm~\ref{algo:construct}). 
The transition rules are applied iteratively. 
When multiple rules are applicable, 
they are applied non-deterministically. 
Pruning terminates once no rule can be applied. 
Let $\tuple{\E_p, \wwcons_p, \wrcons_p}$ be the final state, 
the pruned hyper-polygraph $\G_p$ is $(\V, \E_p, (\wwcons_p, \wrcons_p))$.

\input{figs/prune-rules}

In Figure~\ref{fig:rules-of-pruning}, 
we use $\hookrightarrow$ to denote state transitions, 
distinguishing it from $\rel{}$ (edge) and $\leadsto$ (reachability). 
Pruning relies on two main transitions \textsc{Elim} and \textsc{Intro}.
\textsc{Elim} removes an infeasible choice 
that create cycles in the known graph. 
When a constraint has been reduced to contain a single edge, 
\textsc{Intro} resolves it 
by removing it from $\C$ and 
adding the remaining edge to the known graph.
The prefix $\WW-$ and $\WR-$ of the rule names  
indicate the application of them to $\WW$ and $\WR$ edges, 
respectively.
Pruning may also terminate 
when encountering a violation of the isolation level, 
captured by the rules \textsc{No-Choice} and \textsc{Cycle}.
\textsc{No-Choice} applies when every edge would create a cycle, 
while \textsc{Cycle} applies when the known graph itself is cyclic, 
implying that all compatible graphs are cyclic.
Both cases yield a violation of \ser, 
represented by the special terminal state $\notmodelssymbol$. 

Let $n$ be the number of applied transition rules  
and $s_i$  the $i$-th state, 
the pruning process can be represented as a transition chain:
\[ s_0 \hookrightarrow s_1 \hookrightarrow s_2 \hookrightarrow \cdots \hookrightarrow s_n, \]
where each $s_i$ corresponds to a hyper-polygraph.
Specially, $s_0 \triangleq \tuple{\E, \wwcons, \wrcons}$ corresponds to $\G$
and $s_n \triangleq \tuple{\E_p, \wwcons_p, \wrcons_p} $ corresponds to $\G_p$ 
($s_n$ may be $\notmodelssymbol$).
We write $\tuple{\E, \wwcons, \wrcons} \hookrightarrow^{*} \tuple{\E_p, \wwcons_p, \wrcons_p}$. 
If $\H \models \textsc{SER}$, 
we say that the hyper-polygraph $\G$ of $\H$ is \emph{\textsc{SER}-acyclic}.

\begin{theorem} \label{thm:correctness-of-pruning}
  Pruning is correct, i.e., 
  \begin{enumerate}
    \item If $\tuple{\E, \wwcons, \wrcons} \hookrightarrow^{*} \tuple{\E_p, \wwcons_p, \wrcons_p}$, 
      then $\G$ is \textsc{SER}-acyclic if and only if $\G_p$ is \textsc{SER}-acyclic.
    \item If $\tuple{\E, \wwcons, \wrcons} \hookrightarrow^{*} \notmodelssymbol$, 
      then $\G$ is not \textsc{SER}-acyclic; 
  \end{enumerate}
\end{theorem}

\begin{proof}[Proof of Condition (1)]
  We first prove the condition (1) of Theorem~\ref{thm:correctness-of-pruning} 
  by induction on the number of transitions. 
  In this case, neither rule \textsc{No-Choice} nor \textsc{Cycle} is applied. 
  Let $s_i$ correspond to $\G_i$ for $0 \le i \le n$; in particular,
  $\G_0=\G$ and $\G_n=\G_p$.
  \begin{itemize}
    \item \textbf{Base Case.}
      $s_0$ corresponds to $\G$, which directly satisfies condition (1).
    \item \textbf{Inductive Step.}
      Fix $0 \leq k < n$ and assume $\G_k$ is \textsc{SER}-acyclic if and only if $\G$ is \textsc{SER}-acyclic, 
      we show that $\G_{k + 1}$ is \textsc{SER}-acyclic if and only if $\G_k$ is \textsc{SER}-acyclic.
      Let $s_k \hookrightarrow s_{k+1}$ be obtained by one of the following rules:
      \begin{itemize}
        \item $\WW$-\textsc{Elim} or $\WR$-\textsc{Elim}.
          The removed edge (of type $\WW$ or $\WR$) 
          cannot be present in any acyclic compatible graph of $\G_k$; 
          otherwise, it creates a cycle in the known graph.
          Since the rest of the structure from $\G_k$ to $\G_{k + 1}$ remains unchanged, 
          the sets of acyclic compatible graphs of $\G_k$ and $\G_{k+1}$ are the same.
          Hence $\G_{k+1}$ is \textsc{SER}-acyclic iff $\G_k$ is \textsc{SER}-acyclic.
        \item $\WW$-\textsc{Intro} or $\WR$-\textsc{Intro}.
          By Definition~\ref{def:compatible-graphs-with-a-hyper-polygraph}, 
          when a constraint is reduced to contain a single edge, the remaining edge $e$ 
          (and the $\RW$ edges derived from the known graph) 
          must belong to every compatible graph of $\G_k$.
          Adding these edges to $\E$ merely makes explicit what is
          already required in all compatible graphs; 
          no other part of the structure changes. 
          Therefore the sets of acyclic compatible graphs of $\G_k$ 
          and $\G_{k+1}$ are the same, 
          and again $\G_{k+1}$ is \textsc{SER}-acyclic iff $\G_k$ is \textsc{SER}-acyclic.
      \end{itemize}
  \end{itemize}
  Thus, by induction on $k$, 
  $\G_n=\G_p$ is \textsc{SER}-acyclic iff $\G$ is \textsc{SER}-acyclic.
\end{proof}

\begin{proof}[Proof of Condition (2)]
  As $\notmodelssymbol$ has no outgoing transitions, 
  it is sufficient to show  
  if either the rule \textsc{No-Choice} or \textsc{Cycle} is applied 
  on the last transition $s_{n - 1} \hookrightarrow s_n$, 
  then $\G_{n - 1}$ is not \textsc{SER}-acyclic.
  \begin{itemize}
    \item \textsc{No-Choice}.
      By Definition~\ref{def:compatible-graphs-with-a-hyper-polygraph}, 
      if a constraint $\ccons$ in $\G_{n - 1}$ becomes empty, 
      no compatible graph of $\G_{n - 1}$ exists, 
      since $\cardinality{\E \cap \ccons} = 1$ cannot be satisfied. 
      Hence $\G_{n - 1}$ is not \textsc{SER}-acyclic.
    \item \textsc{Cycle}.
      By Definition~\ref{def:compatible-graphs-with-a-hyper-polygraph}, 
      any compatible graph of $\G_{n - 1}$ must include the cycle already present in the known graph.
      Thus, $\G_{n - 1}$ is not \textsc{SER}-acyclic.
  \end{itemize}
  Hence, in both cases $\G_{n-1}$ is not \textsc{SER}-acyclic. 
  By the equivalence already established between $\G_{n-1}$ and $\G$, 
  it follows that $\G$ is not \textsc{SER}-acyclic.
\end{proof}

\inlsec{Encoding 2-width Cycles}
Let $\G_p$ be the hyper-polygraph after pruning 
and let $\Phi_{\G_p}$ denote its encoding (see Section~\ref{ss:encoding}). 
In the encoding of 2-width cycles, 
$\Phi_{\G_p}$ is extended to the final formula $\F$ by conjoining the encoding of 
all identified 2-width cycles.
Let $n$ be the number of such cycles and 
$\rho_i (1 \leq i \leq n)$ denote the encoding of the i-th cycle.
Formally, 
\[\F = \Phi_{\G_p} \land \bigwedge_{1 \leq i \leq n} \rho_i, 
\text{ where } \rho_i = \lnot v_{i, 1} \lor \lnot v_{i, 2} 
\text{ (see Algorithm~\ref{algo:init-pair-conflict})}. \]

\vspace{-2ex}
\begin{theorem} \label{thm:correctness-of-encode-2-width-cycles}
  Given for the same known graph in the theory solver, 
  $\Phi_{\G_p}$ and $\F$ are equisatisfiable, i.e., 
  $\Phi_{\G_p}$ is satisfiable if and only if $\F$ is satisfiable.
\end{theorem}

\vspace{-2ex}
\begin{proof}
  This proof proceeds by induction on the number $k$ of 2-width cycles.
  Let $\F_k = \Phi_{\G_p} \land \bigwedge\limits_{1 \leq i \leq k} \rho_i$, 
  so that $\F_0 = \Phi_{\G_p}$.  
  \begin{itemize}
    \item \textbf{Base Case.} $\F_0 = \Phi_{\G_p}$ directly implies the equisatisfiability.
    \item \textbf{Inductive Step.} 
      Assume $\F_k$ is equisatisfiable with $\Phi_{\G_p}$ for some $0 \leq k < n$. 
      We show that $\F_{k + 1} = \F_k \land \rho_{k + 1}$ is equisatisfiable with $\F_k$.
      By construction, $\rho_{k+1} = \lnot v_{k+1,1} \lor \lnot v_{k+1,2}$ encodes the
      fact that the two edges associated with $v_{k+1,1}$ and $v_{k+1,2}$ cannot 
      simultaneously exist, otherwise they would form a 2-width cycle in the known
      graph (see Section~\ref{sss:encode-2-width-cycles}). 
      Therefore \emph{every} satisfying model of $\F_{k}$ satisfies $\rho_{k+1}$. 
      Hence $\F_{k}$ is equisatisfiable with $\F_{k} \land \rho_{k+1}$, and by the
      definition of $\F_{k+1}$ we obtain that $\F_{k+1}$ is equisatisfiable with $\F_{k}$.
  \end{itemize}
\end{proof}

\subsection{Order-Guided Polarity Picking}
\label{ss:appendix-correctness-polarity}

\begin{theorem} \label{thm:correctness-of-polarity}
  The polarity picking heuristic described in Section~\ref{ss:polarity} 
  does not affect the correctness of the CDCL(T) algorithm.
\end{theorem}

\begin{proofsketch}
  Let $\cdclt$ denote the CDCL(T) algorithm without the polarity picking heuristic ($\textbf{H}$), 
  and let $\cdclth$ denote the algorithm with the heuristic $\mathbf{H}$ applied.
  The heuristic $\mathbf{H}$ neither introduces new branches 
  nor eliminates existing ones in the search. 
  Consequently, $\cdclt$ and $\cdclth$ explore the same search space, 
  possibly in different orders.  
  Hence, the correctness of $\mathbf{H}$ follows directly 
  from the correctness of  CDCL(T)~\cite{SMT:JACM2006}.
\end{proofsketch}

%% file: figs/prune-rules.tex

\begin{figure*}[t]  
\scriptsize
  \centering
  \begin{mathpar}
    \large
    \makebox[\linewidth][c]{%
      \begin{minipage}[t]{0.46\linewidth}\centering
      \(\inferrule*[right=\large\textsc{No-Choice}]
        { \emptyset \in \wwcons \cup \wrcons }
        { \tuple{\E, \wwcons, \wrcons} \hookrightarrow \notmodelssymbol } \)
      \end{minipage}
      \hspace{0.06\linewidth} 
      \begin{minipage}[t]{0.46\linewidth}\centering
      \(\inferrule*[right=\large\textsc{Cycle}]
        { \E \text{ is cyclic } }
        { \tuple{\E, \wwcons, \wrcons} \hookrightarrow \notmodelssymbol } \)
      \end{minipage}
    }

    \inferrule*[right=\large$\WW$-\textsc{Elim}]
    { \ccons \in \wwcons \and 
      T \rel{\WW(x)} T' \in \ccons \and
      \ccons' \coloneqq \ccons \setminus \set{T \rel{\WW(x)} T'} \\
      \E^{\WW} \coloneqq \set{T \rel{\WW(x)} T'} \cup \set{S \rel{\RW(x)} T' \mid T \rel{\WR(x)} S \in \E} \and 
      \E \cup \E^{\WW} \text{ is cyclic } }
    { \tuple{\E, \wwcons, \wrcons}
      \hookrightarrow
      \tuple{\E, (\wwcons \setminus \set{\ccons}) \cup \set{\ccons'}, \wrcons } }

    \inferrule*[right=\large$\WR$-\textsc{Elim}]
    { \ccons \in \wrcons \and 
      T \rel{\WR(x)} S \in \ccons \and
      \ccons' \coloneqq \ccons \setminus \set{T \rel{\WR(x)} S} \\ 
      \E^{\WR} \coloneqq \set{T \rel{\WR(x)} S} \cup \set{S \rel{\RW(x)} T' \mid T \rel{\WW(x)} T' \in \E} \and 
      \E \cup \E^{\WR} \text{ is cyclic } }
    { \tuple{\E, \wwcons, \wrcons}
      \hookrightarrow
      \tuple{\E, \wwcons, (\wrcons \setminus \set{\ccons}) \cup \set{\ccons'}} }

    \inferrule*[right=\large$\WW$-\textsc{Intro}]
    { \set{T \rel{\WW(x)} T'} \in \wwcons \\ \E^{\WW} \coloneqq \set{T \rel{\WW(x)} T'} \cup \set{S \rel{\RW(x)} T' \mid T \rel{\WR(x)} S \in \E} }
    { \tuple{\E, \wwcons, \wrcons}
      \hookrightarrow
      \tuple{\E \cup \E^{\WW}, \wwcons \setminus \set{T \rel{\WW(x)} T'}, \wrcons} }

    \inferrule*[right=\large$\WR$-\textsc{Intro}]
    { \set{T \rel{\WR(x)} S} \in \wrcons \\ \E^{\WR} \coloneqq \set{T \rel{\WR(x)} S} \cup \set{S \rel{\RW(x)} T' \mid T \rel{\WW(x)} T' \in \E} }
    { \tuple{\E, \wwcons, \wrcons}
      \hookrightarrow
      \tuple{\E \cup \E^{\WR}, \wwcons, \wrcons \setminus \set{T \rel{\WR(x)} S}} }
    


  \end{mathpar}
\captionsetup{skip=2pt}
  \caption{Rules for eliminating 1-width cycles.}
  \label{fig:rules-of-pruning}

\end{figure*}

%% file: appendix/app-eval.tex

\section{Memory Usage Comparison}
\label{section:appendix-memory}

\input{figs/eval/various/mem.tex}

Figure~\ref{fig:various-memory} shows the peak memory usage of \ourtool and the baseline on various general workloads, 
complementing Figure~\ref{fig:various-runtime}; see Section~\ref{subsubsec:vs-baseline}.

\input{tables/notations}

%% file: figs/eval/various/mem.tex

\pgfplotsset{height=140pt, width=200pt}

\begin{figure}[t]
	\centering
	\resizebox{\linewidth}{!}{
	\begin{scaletikzpicturetowidth}{0.3\textwidth}
		\begin{tikzpicture}[scale=\tikzscale]
			\begin{axis}[
				font=\Large,
				title={(a)},
				xlabel={\#sessions},
				ylabel={Memory (MB)},
				ymax=600,
				cycle multiindex* list={
								color       \nextlist
								mark list*  \nextlist
						},
				legend pos=north west,
				 legend style={draw=none}
			]
				\addplot[color=blue,mark=o,mark size=3pt] table [x=param, y=ours, col sep=comma] {figs/eval/various/data/mem-sessions.csv};
				\addplot[color=brown,mark=square,mark size=3pt] table [x=param, y=baseline-pruning, col sep=comma] {figs/eval/various/data/mem-sessions.csv};
				\addplot[color=violet,mark=triangle,mark size=3pt] table [x=param, y=baseline, col sep=comma] {figs/eval/various/data/mem-sessions.csv};
				\legend{\ourtool, baseline+P, baseline}
			\end{axis}
		\end{tikzpicture}
		\hspace{1ex}
	\end{scaletikzpicturetowidth}
	\begin{scaletikzpicturetowidth}{0.27\textwidth}
		\begin{tikzpicture}[scale=\tikzscale]
			\begin{axis}[
				font=\Large,
				title={(b)},
				xlabel={\#txns/session},
				ymax=1000,
				legend pos=north west,
			]
				\addplot[color=blue,mark=o,mark size=3pt] table [x=param, y=ours, col sep=comma] {figs/eval/various/data/mem-txns.csv};
				\addplot[color=brown,mark=square,mark size=3pt] table [x=param, y=baseline-pruning, col sep=comma] {figs/eval/various/data/mem-txns.csv};
				\addplot[color=violet,mark=triangle,mark size=3pt] table [x=param, y=baseline, col sep=comma] {figs/eval/various/data/mem-txns.csv};
			\end{axis}
		\end{tikzpicture}
				\hspace{1ex}
	\end{scaletikzpicturetowidth}
	
	\begin{scaletikzpicturetowidth}{0.27\textwidth}
		\begin{tikzpicture}[scale=\tikzscale]
			\begin{axis}[
				font=\Large,
				title={(c)},
				xlabel={\#ops/txn},
				ymax=1000,
				legend pos=north west,
			]
      \addplot[color=blue,mark=o,mark size=3pt] table [x=param, y=ours, col sep=comma] {figs/eval/various/data/mem-evts.csv};
      \addplot[color=brown,mark=square,mark size=3pt] table [x=param, y=baseline-pruning, col sep=comma] {figs/eval/various/data/mem-evts.csv};
			\addplot[color=violet,mark=triangle,mark size=3pt] table [x=param, y=baseline, col sep=comma] {figs/eval/various/data/mem-evts.csv};
  		\end{axis}
		\end{tikzpicture}
	\end{scaletikzpicturetowidth}
	}
\vspace{1ex}
	\resizebox{\linewidth}{!}{

	\begin{scaletikzpicturetowidth}{0.3\textwidth}
		\begin{tikzpicture}[scale=\tikzscale]
			\begin{axis}[
				font=\Large,
				title={(d)},
				xlabel={\#keys},
				 ylabel={Memory (MB)},
				xtick={2000,4000,6000,8000,10000},
				xticklabels={2k,4k,6k,8k,10k},
				ymax=500,
				legend pos=north west,
				scaled ticks=false
			]
      \addplot[color=blue,mark=o,mark size=3pt] table [x=param, y=ours, col sep=comma] {figs/eval/various/data/mem-keys.csv};
      \addplot[color=brown,mark=square,mark size=3pt] table [x=param, y=baseline-pruning, col sep=comma] {figs/eval/various/data/mem-keys.csv};
			\addplot[color=violet,mark=triangle,mark size=3pt] table [x=param, y=baseline, col sep=comma] {figs/eval/various/data/mem-keys.csv};
    \end{axis}
		\end{tikzpicture}
		\hspace{1ex}
	\end{scaletikzpicturetowidth}

	\begin{scaletikzpicturetowidth}{0.27\textwidth}
		\begin{tikzpicture}[scale=\tikzscale]
			\begin{axis}[
				font=\Large,
				title={(e)},
				xlabel={read proportion\%},
				xtick={5,25,50,75,95},
				ymax=500,
				legend pos=north west,
			]
      \addplot[color=blue,mark=o,mark size=3pt] table [x=param, y=ours, col sep=comma] {figs/eval/various/data/mem-readpct.csv};
      \addplot[color=brown,mark=square,mark size=3pt] table [x=param, y=baseline-pruning, col sep=comma] {figs/eval/various/data/mem-readpct.csv};
			\addplot[color=violet,mark=triangle,mark size=3pt] table [x=param, y=baseline, col sep=comma] {figs/eval/various/data/mem-readpct.csv};
    \end{axis}
		\end{tikzpicture}
				\hspace{1ex}
	\end{scaletikzpicturetowidth}

	\begin{scaletikzpicturetowidth}{0.27\textwidth}
		\begin{tikzpicture}[scale=\tikzscale]
			\begin{axis}[
				font=\Large,
				title={(f)},
				xlabel={dup key proportion\%},
				xtick={0,20,40,60,80,100},
				ymax=300,
				legend pos=north west,
			]
      \addplot[color=blue,mark=o,mark size=3pt] table [x=param, y=ours, col sep=comma] {figs/eval/various/data/mem-dupkeypct.csv};
      \addplot[color=brown,mark=square,mark size=3pt] table [x=param, y=baseline-pruning, col sep=comma] {figs/eval/various/data/mem-dupkeypct.csv};
			\addplot[color=violet,mark=triangle,mark size=3pt] table [x=param, y=baseline, col sep=comma] {figs/eval/various/data/mem-dupkeypct.csv};
    \end{axis}
		\end{tikzpicture}
	\end{scaletikzpicturetowidth}
	}
	\captionsetup{skip=0pt}
  \caption{
		Peak memory usage comparison.
		Timeout data points are not plotted.
	}
	\label{fig:various-memory}
	\vspace{-2ex}
\end{figure}
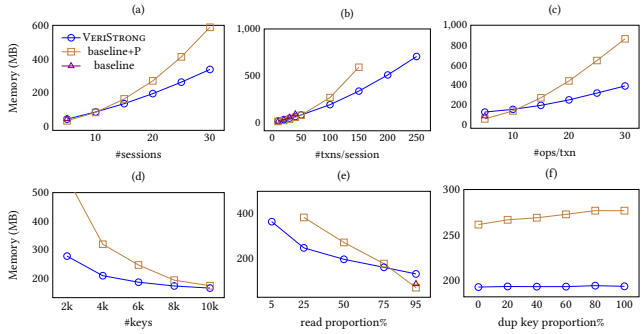

%% file: tables/notations.tex

\begin{table*}[t]
\footnotesize
	\centering
		\captionsetup{skip=5pt}
	\caption{Notations}
	\label{table:notations}
	\begin{tabular}{|c|c|c|}
	\hline
	\textbf{Category}                   & \textbf{Notation}          & \textbf{Description}                  \\ \hline
	\multirow{3}{*}{\textit{KV Store}}  & $\Key$                     & set of keys                   \\ \cline{2-3}
										& $\Val$                     & set of values                 \\ \cline{2-3}
										& $\Op$                      & set of operations             \\ \hline
  \textit{Dependency}                 & $\SO$, $\WR$, $\WW$, $\RW$ & dependency relations/edges        \\ \hline
	\textit{History}                    & $\H = (\T, \SO)$ & history        \\ \hline
	\multirow{4}{*}{\textit{Hyper-polygraph}}     & $\G = (\V, \E, \C)$            & hyper-polygraph           \\ \cline{2-3}
										& $\C = (\wwcons, \wrcons)$                               & constraint \\ \cline{2-3}
										& $\G' = (\V', \E')$                               & compatible graph \\ \cline{2-3}
										& $T \rel{\SO} S, T \rel{\WW(x)} S, T \rel{\WR(x)} S, T \rel{\RW(x)} S$                   & edges \\ \hline
  \multirow{5}{*}{\textit{Propositional Logic}}     & $v$            & variable           \\ \cline{2-3}
										& $l = v$ or $\lnot v$  & literal \\ \cline{2-3}
										& $C = l_1 \lor l_2 \lor \cdots \lor l_n$  & clause \\ \cline{2-3}
										& $\formula = C_1 \land C_2 \land \cdots \land C_m$  & formula \\ \cline{2-3}
										& $\pa$  & (partial) model \\ \hline
  \multirow{5}{*}{\textit{SMT}}     & $\csovar_{i, j}, \cwwvar_{i, j}^x, \cwrvar{x}_{i, j}, \crwvar_{i, j}^x$            & variables corresponding to different edges         \\ \cline{2-3}
										& $\Phi_{\G} = \Phi_{\mathit{KG}} \land \Phi_{\wwcons} \land \Phi_{\wrcons} \land \Phi_{\RW}$  & encodings \\ \cline{2-3}
										& $\cvars$  & set of boolean variables \\ \cline{2-3}
										& $\cclauses$  & set of clauses \\ \cline{2-3}
										& $\ug = (\V, \eug)$  & the graph maintained in the theory solver \\ \hline
	\end{tabular}%
	\end{table*}

%% file: appendix/app-enea.tex

\section{Characterizing Biswas et al.'s Theory}
\label{section:appendix-enea}

In the theory of Biswas et al.~\cite{Complexity:OOPSLA2019},
a history satisfies a given isolation level
if there exists a strict total order $\CO$ (called \emph{commit order})
on its transactions that extends both the $\WR$ and $\SO$ relations
and satisfies certain additional properties.
In this section, we show how to characterize isolation levels
under this formalism using hyper-polygraphs,
taking \ser{} as a representative example.


\ser{} requires that for any transaction $T_1$
writing to a key $x$ that is read by a transaction $T_3$,
every commit predecessor of $T_3$ that also writes to $x$
must precede $T_1$ in commit order. That is,

\begin{theorem} 
	\label{thm:ser-commit-order}
	For a history \emph{$\H = (\T, \SO)$},
	\emph{\begin{align*}
		\H \models\,\, &  \textsc{SER}  \iff \H \models \intaxiom \;\land
			 \exists \CO.\; \forall x \in \Key.\;
       \forall T_1, T_2, T_3 \in \T.\; \\
      &T_1 \neq T_2 \land
       T_1 \rel{\WR(x)} T_3 \land
       T_2 \vdash \writeevent(x, \_)\;\land \\
      &(T_2, T_3) \in \CO \implies
       (T_2, T_1) \in \CO.
        \end{align*}}
\end{theorem}

The definition of hyper-polygraph under Biswas et al.'s formalism
is a straightforward adaptation of Definition~\ref{def:hyper-polygraph},
with the constraint set $\wwcons$ replaced by $\cocons$.

\begin{definition} \label{def:hyper-polygraph-enea}
  A hyper-polygraph $\G = (\V, \E, \C)$
  for a history $\H = (\T, \SO)$
  is a directed labeled graph $(\V, \E)$, referred to as the known graph,
  together with \emph{a pair of constraint sets} $\C = (\cocons, \wrcons)$, where
  \begin{itemize}[leftmargin=15pt]
    \item $\V$ is the set of nodes, corresponding to the transactions in $\H$;
    \item $\E \subseteq \V \times \V \times \type \times \Key$
      is a set of edges,
      where each edge is labeled with a dependency type from
      $\type = \set{\SO, \WR, \CO}$ and a key from $\Key$;
    \item $\cocons$ is a constraint set over  uncertain version orders, defined as
      $\cocons = \Bset{\btuple{T \rel{\CO} S, S \rel{\CO} T}
      \mid T, S \in \T \land T \neq S}$; and
    \item $\wrcons$ is a constraint set over  uncertain read-from mappings, defined as
      $\wrcons = \Bset{\bigcup\limits_{T_{i} \vdash \writeevent(x, v)} \set{T_{i} \rel{\WR(x)} S}
      \mid S \vdash \readevent(x, v)}$.
  \end{itemize}
\end{definition}

A compatible graph of the  hyper-polygraph
resolves the uncertainty in $\CO$ and $\WR$,
while enforcing the $\CO$-required properties.


\begin{definition} \label{def:compatible-graphs-with-a-hyper-polygraph-of-ser-enea}
  A directed labeled graph $\G' = (\V', \E')$ is said to be \emph{$\textsc{SER}$-compatible} with
  a hyper-polygraph $\G = (\V, \E, \C)$ if
  \begin{itemize}[leftmargin=25pt]
    \item $\V' = \V$;
    \item $\E' \supseteq \E$ such that
      $\forall T_1, T_2, T_3 \in \V.\;
       T_1 \neq T_2 \land
       T_1 \rel{\WR(x)} T_3 \land
       T_2 \in \WriteTx_x \land
       T_2 \rel{\CO} T_3 \implies
       T_2 \rel{\CO} T_1$;
    \item $\forall \consvar \in \cocons.\;
      |\E' \cap \consvar| = 1$; and
    \item $\forall \consvar \in \wrcons.\;
      |\E' \cap \consvar| = 1$.
  \end{itemize}
\end{definition}

Consequently, the hyper-polygraph-based characterization of \ser{}
in Biswas et al.'s formalism follows directly from
Theorem~\ref{thm:ser-commit-order}.

\begin{theorem}
  \label{thm:ser-hpg-enea}
  A history $\H$ satisfies \emph{\ser{}} if and only if
  \emph{$\H \models \intaxiom$} and
  there exists an acyclic graph that is $\textsc{SER}$-compatible with
  the hyper-polygraph of $\H$.
\end{theorem}

Similarly, other isolation guarantees defined in Biswas et al.'s framework
can also be characterized using hyper-polygraphs.
Note that \rc requires a finer-grained modeling of the program order ($\po$).




%% file: appendix/app-notation-table.tex

\section{Notations}
\label{section:appendix-notations}

Table~\ref{table:notations} provides a summary of the notations used in this paper.